\documentclass[fontsize=9pt,DIV=calc,a4paper,twocolumn]{scrartcl}

\usepackage[T1]{fontenc}
\usepackage[format=plain]{caption}
\usepackage{microtype,amsmath,amssymb,mathtools,physics,textcomp,booktabs,siunitx,multirow,array,xtab,enumitem,cuted,upgreek}
\usepackage[a4paper,left=15mm,right=15mm,top=15mm,bottom=20mm]{geometry}
\usepackage[auth-lg,affil-it]{authblk}
\usepackage[sort&compress,numbers]{natbib}
\usepackage[hidelinks]{hyperref}
\setlist{noitemsep,topsep=0pt,parsep=0pt,partopsep=0pt,leftmargin=*}

\setkomafont{disposition}{\normalsize\mdseries\bfseries}
\setkomafont{subsection}{\normalsize\mdseries\itshape}

\DeclareMathOperator{\Div}{div}

\def\tabscale{0.82}

\title{Experimental parameter uncertainty in PEM fuel cell modeling\\Part I: Scatter in material parameterization}

\begin{document}

\author{Roman Vetter}
\author{J\"{u}rgen O.~Schumacher}
\affil{Institute of Computational Physics (ICP),\\Zurich University of Applied Sciences (ZHAW),\\Wildbachstrasse 21, CH-8401 Winterthur, Switzerland}

\twocolumn[
\begin{@twocolumnfalse}
\maketitle
\vskip1\baselineskip
\begin{abstract}
Ever since modeling has become a mature part of proton exchange membrane fuel cell (PEMFC) research and development, it has been plagued by significant uncertainty lying in the detailed knowledge of material properties required. Experimental data published on several transport coefficients are scattered over orders of magnitude, even for the most extensively studied materials such as Nafion membranes, for instance. For PEMFC performance models to become predictive, high-quality input data is essential. In this bipartite paper series, we determine the most critical transport parameters for which accurate experimental characterization is required in order to enable performance prediction with sufficient confidence from small to large current densities. In the first part, a macro-homogeneous two-phase membrane-electrode assembly model is furnished with a comprehensive set of material parameterizations from the experimental and modeling literature. The computational model is applied to demonstrate the large spread in performance prediction resulting from experimentally measured or validated material parameterizations alone. The result of this is a ranking list of material properties, sorted by induced spread in the fuel cell performance curve. The three most influential parameters in this list stem from membrane properties: The Fickean diffusivity of dissolved water, the protonic conductivity and the electro-osmotic drag coefficient.
\end{abstract}
\vskip3\baselineskip
\end{@twocolumnfalse}]{}

\section{Introduction}

Fuel cell researchers that consider numerical modeling for their work are confronted with the difficult question which material properties to measure and plug into the models in order to obtain reliable simulation results. With finite financial and time budgets at hand, effort put into the characterization of membrane-electrode assembly (MEA) components must be prioritized. Is it worthwhile to determine the tortuosity of gas diffusion pathways through the porous layers with high resolution imaging, or should the focus be on precise measurements of water transport through the proton exchange membrane (PEM)? How will uncertainties in these parameters affect the predictive power of a detailed fuel cell model? MEA engineers face similar problems: For the efficient development of improved multifunctional materials for PEM fuel cells (PEMFCs), it is critical to know what quantity to optimize for. Will a thinner membrane yield significant improvements in fuel cell performance, or is it more productive to look into reducing contact resistances? Even when the MEA setup is fixed and known in detail, modelers are confronted with the problem of picking the right material parameterization among the many that have been proposed in the open literature. Published data on the most critical transport processes occurring in PEM fuel cells (\textit{NB} for the very same materials) are sometimes scattered significantly, which raises the question which to adopt.

Recently, we have demonstrated how difficult it can be to reliably predict fuel cell performance with numerical modeling under the present spread of experimental material data in the literature \cite{vetter:18b}. The water diffusivity in the membrane \cite{kusoglu:12b,kusoglu:17}, electro-osmotic drag \cite{dai:09}, protonic conductivity of the membrane \cite{liu:16} and the phase change rates of water \cite{marek:01} are some out of several properties that are required for state-of-the-art two-phase PEMFC models and for which measured data disagree largely; further ones are discussed in this paper.

Not many studies systematically address the variation in performance prediction of fuel cell models arising from uncertainty in their parameterization. Typically, a single specific constitutive \emph{parameterization} is implemented for each material property, consisting of a functional relationship and a set of fit \emph{parameters}. Only the latter have been varied to estimate the model sensitivity to them (e.g., \cite{min:06,zhao:15,laoun:16}). Uncertainty resulting from the choice of the complete constitutive relationships, on the other hand, is largely unexplored. In Part I of this bipartite paper series, we intend to fill this gap. After a brief model summary in Sec.~\ref{sec:model}, we discuss its constitutive parameterization in detail in Sec.~\ref{sec:parameterization}, reviewing the literature on several essential material properties. With this comprehensive database of proposed material parameterizations at hand, the scatter in fuel cell performance resulting from it will be presented in Sec.~\ref{sec:results}. This allows us to conclude with a sorted list of MEA properties which cause uncertainty in the resulting polarization curve and for which more reliable experimental characterization is therefore required. In Part II \cite{vetter:p2}, we will extend this survey by carrying out extensive forward uncertainty propagation analyses to get a more rigorous insight into the relative importance of PEMFC material parameterizations.

\section{Fuel cell model}
\label{sec:model}

In order to get a quantitative picture of the uncertainty in the performance prediction of state-of-the-art PEMFC models due to disagreement or scatter in the experimental literature data on fuel cell materials and processes, a model is needed that is both easily modifiable and numerically efficient. Here, we build upon our previously developed macro-homogeneous, steady-state, two-phase model of a five-layer MEA \cite{vetter:18}. Being built for easy parameter substitution and fast execution of many different simulations (e.g., with different constitutive material properties), this one-dimensional model offers the best middle ground between model complexity, extensibility and computational efficiency. It models the coupled charge, heat and mass transport processes in through-plane direction within a differential PEMFC. For improved accuracy, we extend it here in a number of ways: Among other amendments (detailed further below), multicomponent gas diffusion is accounted for with the Maxwell--Stefan equations, Knudsen diffusion is included, thermo-osmosis and Schroeder's paradox are incorporated, and electrical as well as thermal contact resistance between the MEA layers are added to the model. In this section we briefly recapitulate the governing equations and refer to \cite{vetter:18} for additional information.

\subsection{Conservation laws}

The MEA modeled here consist of a classical symmetrical arrangement of five porous layers: A polymer electrolyte membrane (PEM, Nafion) in the middle, sandwiched by two thin catalyst layers (ACL \& CCL) containing an electron-conducting carbon matrix and a proton-conducting ionomer phase (Nafion), which in turn are clamped between two fibrous gas diffusion layers made of hydrophobized carbon paper (AGDL \& CGDL).

For the conservation of electrons (subscript e) and protons (p), Ohm's law
\begin{equation}
\begin{aligned}
j_\mathrm{e} = -\sigma_\mathrm{e} \nabla \varphi_\mathrm{e}, \quad \Div j_\mathrm{e} &= S_\mathrm{e}\\
j_\mathrm{p} = -\sigma_\mathrm{p} \nabla \varphi_\mathrm{p}, \quad \Div j_\mathrm{p} &= S_\mathrm{p}\\
\end{aligned}
\end{equation}
is solved with effective layer-dependent conductivities $\sigma_\mathrm{e}$ and $\sigma_\mathrm{p}$ (to be specified in Sec.~\ref{sec:ionic_conductivity}). The source terms read
\begin{equation}
S_\mathrm{e} =
\begin{dcases*}
0 & in AGDL\\
-S_\mathrm{A} & in ACL\\
S_\mathrm{C} & in CCL\\
0 & in CGDL
\end{dcases*},\quad
S_\mathrm{p} =
\begin{dcases*}
S_\mathrm{A} & in ACL\\
0 & in PEM\\
-S_\mathrm{C} & in CCL
\end{dcases*}
\end{equation}
where the anode and cathode reaction rates $S_\mathrm{A}$ and $S_\mathrm{C}$ will be specified in Sec.~\ref{sec:reaction}. In the CLs, the two phase potentials $\varphi_\mathrm{e}$ and $\varphi_\mathrm{p}$ coexist, defining the total Galvani potential difference $\Delta\varphi=\varphi_\mathrm{e}-\varphi_\mathrm{p}$. Using the reversible half-cell potentials
\begin{equation}
\begin{split}
\Delta\varphi^0_\mathrm{A} &= -\frac{T\Delta S^\mathrm{ref}_\mathrm{A}}{2F} - \frac{RT}{2F}\ln\left[\frac{p_{\mathrm{H}_2}}{P_\mathrm{ref}}\right]\\
\Delta\varphi^0_\mathrm{C} &= -\frac{\Delta H^\mathrm{ref}-T\Delta S^\mathrm{ref}_\mathrm{C}}{2F} + \frac{RT}{4F}\ln\left[\frac{p_{\mathrm{O}_2}}{P_\mathrm{ref}}\right]
\end{split}
\end{equation}
the anode and cathode activation overpotentials are defined as
\begin{equation}
\eta_\mathrm{A} = \Delta\varphi - \Delta\varphi^0_\mathrm{A}\quad\text{and}\quad\eta_\mathrm{C} = \Delta\varphi^0_\mathrm{C} - \Delta\varphi.
\end{equation}
$\Delta H^\mathrm{ref}$ is the standard enthalpy of formation of liquid water, $\Delta S_\mathrm{A}^\mathrm{ref}$ and $\Delta S_\mathrm{C}^\mathrm{ref}$ the half-reaction entropies, $F$ Faraday's constant, $R$ the gas constant, $T$ the absolute temperature, $P_\mathrm{ref}$ the reference pressure, $p_{\mathrm{H}_2}=y_{\mathrm{H}_2}P$ the partial pressure of hydrogen, and $p_{\mathrm{O}_2}=y_{\mathrm{O}_2}P$ that of oxygen.

For the conservation of energy, heat conduction is considered according to Fourier's law across all layers:
\begin{equation}
j_T = -k \nabla T, \quad \Div j_T = S_T
\end{equation}
with effective thermal conductivity $k$ (Sec.~\ref{sec:thermal_conductivity}). The total heat source $S_T$ includes the following contributions: Ohmic heating in the electron and proton-conducting phases, Peltier heating and thermal activation losses, and the latent heat of water ab-/desorption (subscript ad) in the CLs and evaporation/condensation (ec) where liquid water is present (assumed only on the cathode side of the membrane). This yields
\begin{strip}
\hrule\vskip3pt
\begin{equation}
\label{eq:ST}
S_T =\begin{dcases*}
j_\mathrm{e}^2/\sigma_\mathrm{e} \phantom{{}+ j_\mathrm{p}^2/\sigma_\mathrm{p} + (\Pi_\mathrm{A} + \eta_\mathrm{A})S_\mathrm{A} +H_\mathrm{ad}S_\mathrm{ad} + H_\mathrm{ec}S_\mathrm{ec}} & in AGDL\\
j_\mathrm{e}^2/\sigma_\mathrm{e} + j_\mathrm{p}^2/\sigma_\mathrm{p} + (\Pi_\mathrm{A} + \eta_\mathrm{A})S_\mathrm{A} + H_\mathrm{ad}S_\mathrm{ad} \phantom{{}+ H_\mathrm{ec}S_\mathrm{ec}} & in ACL\\
\phantom{j_\mathrm{e}^2/\sigma_\mathrm{e} +{}}j_\mathrm{p}^2/\sigma_\mathrm{p} & in PEM\\
j_\mathrm{e}^2/\sigma_\mathrm{e} + j_\mathrm{p}^2/\sigma_\mathrm{p} + (\Pi_\mathrm{C} + \eta_\mathrm{C})S_\mathrm{C} + H_\mathrm{ad}S_\mathrm{ad} + H_\mathrm{ec}S_\mathrm{ec} & in CCL\\
j_\mathrm{e}^2/\sigma_\mathrm{e} \phantom{{}+ j_\mathrm{p}^2/\sigma_\mathrm{p} + (\Pi_\mathrm{C} + \eta_\mathrm{C})S_\mathrm{C} + H_\mathrm{ad}S_\mathrm{ad}} + H_\mathrm{ec}S_\mathrm{ec} & in CGDL
\end{dcases*}
\end{equation}
\hrule
\end{strip}
where $\Pi_\mathrm{A,C}=-\Delta S_\mathrm{A,C}^\mathrm{ref}T/2F$ are the Peltier coefficients of the half-reactions, $S_\mathrm{ad}$ ($S_\mathrm{ec}$) the reaction rate of water ab-/desorption (evaporation/condensation) detailed in Sec.~\ref{sec:phase_change}, and $H_\mathrm{ad}$ ($H_\mathrm{ec}$) the corresponding latent heat (Sec.~\ref{sec:latent_heat}).

The transport of dissolved water through the membrane and the ionomer phase of the CLs is modeled in the spirit of Springer et al.~\cite{springer:91}, who simplified the general diffusion equation with chemical potential gradient into Fick's law combined with electro-osmotic drag. Here we additionally account for thermo-osmosis by writing
\begin{equation}
\label{eq:j_lambda}
j_\lambda = - \frac{D_\lambda}{V_\mathrm{m}}\nabla\lambda + \frac{\xi}{F}j_\mathrm{p} - D_T\nabla T, \quad \Div j_\lambda = S_\lambda
\end{equation}
where $\lambda$ is the hydration number of the ionomer, $V_\mathrm{m}$ its dry equivalent volume (Sec.~\ref{sec:water_diff}), $D_\lambda$ the effective water diffusivity (Sec.~\ref{sec:water_diff}), $\xi$ the electro-osmotic drag coefficient (Sec.~\ref{sec:electro_osmosis}), and $D_T$ the thermo-osmotic transport coefficient (Sec.~\ref{sec:thermo_osmosis}). The source term for dissolved water reads
\begin{equation}
\label{eq:Slambda}
S_\lambda =
\begin{dcases*}
S_\mathrm{ad} & in ACL\\
0 & in PEM\\
S_\mathrm{ad} + \omega S_\mathrm{C}/2F & in CCL
\end{dcases*}.
\end{equation}
$\omega\in[0,1]$ determines the mass fraction of water produced in dissolved rather than in liquid form at the ionomer--catalyst--carbon triple phase boundary. Since the appropriate value for $\omega$ is an open problem \cite{wu:10,cao:13}, we assume $\omega=1/2$. The other half will appear as a source term for liquid water in Eq.~\ref{eq:Ss}.

Assuming that (i) gas convection is negligible everywhere ($\nabla P\equiv0$), (ii) no gas crossover through the membrane occurs, and that (iii) the supplied gases only consist of hydrogen on the anode side and an oxygen-nitrogen mix on the cathode side when dry, the total gas pressure $P$ can be evaluated on these respective sides as
\begin{equation}
P = \begin{cases*}
P_\mathrm{A}=p_{\mathrm{H}_2} + p_{\mathrm{H}_2\mathrm{O}} & in AGDL \& ACL\\
P_\mathrm{C}=p_{\mathrm{O}_2} + p_{\mathrm{H}_2\mathrm{O}} + p_{\mathrm{N}_2} & in CCL \& CGDL
\end{cases*}.
\end{equation}
The ideal gas law is assumed to hold, such that the partial pressures are given by $p_X = y_XCRT$, $X=\mathrm{H}_2,\mathrm{O}_2,\mathrm{H}_2\mathrm{O},\mathrm{N}_2$, where $y_X$ are the mole fractions of the gas species and $C$ is the total interstitial gas concentration. Under these conditions and further assuming that thermal diffusion is negligible, an appropriate way to model the transport of gas species in PEMFC diffusion media is given by an in-series combination of the transport resistances of the Maxwell--Stefan model and Knudsen diffusion:
\begin{equation}
\label{eq:maxwell_stefan}
-C\nabla y_X = \sum_{Y\neq X}\frac{y_Yj_X-y_Xj_Y}{\mathcal{D}_{X,Y}} + \frac{j_X}{D_{\mathrm{K},X}}
\end{equation}
where $\mathcal{D}_{X,Y}$ denote the effective binary diffusivities and $D_{\mathrm{K},X}$ the effective Knudsen diffusivities in the porous layers (Sec.~\ref{sec:gas_diff}). The inert nitrogen is not explicitly modeled on the cathode side because it follows from the requirement that $\sum_Xy_X=1$. Species conservation is imposed with
\begin{equation}
\Div j_X = S_X,\qquad X=\mathrm{H}_2, \mathrm{O}_2, \mathrm{H}_2\mathrm{O}
\end{equation}
where the source terms
\begin{equation}
S_{\mathrm{H}_2} = 
\begin{dcases*}
0 & in AGDL\\
-S_\mathrm{A}/2F & in ACL
\end{dcases*}
\end{equation}
\begin{equation}
S_{\mathrm{O}_2} = 
\begin{dcases*}
-S_\mathrm{C}/4F & in CCL\\
0 & in CGDL
\end{dcases*}
\end{equation}
account for reactant consumption, whereas
\begin{equation}
\label{eq:SH2O}
S_{\mathrm{H}_2\mathrm{O}} =
\begin{dcases*}
0 & in AGDL\\
-S_\mathrm{ad} & in ACL\\
-S_\mathrm{ec} - S_\mathrm{ad} & in CCL\\
-S_\mathrm{ec} & in CGDL
\end{dcases*}
\end{equation}
accounts for the phase transitions from vapor to liquid or dissolved water and back. Liquid water transport is modeled with Darcy's law with the pore saturation $s$ as the dependent variable, i.e,
\begin{equation}
\label{eq:j_s}
j_s = -\frac{D_s}{V_\mathrm{w}}\nabla s,\quad D_s = \frac{K_\mathrm{abs}K_\mathrm{rel}}{\mu} \frac{\partial p_\mathrm{c}}{\partial s},\quad \Div j_s = S_s
\end{equation}
where $V_\mathrm{w}$ is the molar volume of liquid water, $K_\mathrm{abs}$ the porous medium's intrinsic (absolute) hydraulic permeability, $K_\mathrm{rel}$ the saturation-dependent relative permeability, $\partial p_\mathrm{c}/\partial s$ the layer's differential relationship between capillary pressure and saturation, and $\mu$ the dynamic viscosity of liquid water (all detailed in Sec.~\ref{sec:liquid_water}). To balance the evaporation/condensation sinks in Eq.~\ref{eq:SH2O} and to complete the electrochemical production of water halfway accounted for with Eq.~\ref{eq:Slambda}, the liquid water source term must read
\begin{equation}
\label{eq:Ss}
S_s =
\begin{dcases*}
S_\mathrm{ec} + (1-\omega)S_\mathrm{C}/2F & in CCL\\
S_\mathrm{ec} & in CGDL
\end{dcases*}.
\end{equation}

\subsection{Phase change}
\label{sec:phase_change}

Within the CLs, water is absorbed by the ionomer phase and desorbed again back to vapor in a relatively sluggish process (see Sec.~\ref{sec:vapor_sorption}). In order to account for this interfacial mass transfer resistance across the entire CL thickness, the sorption source terms in Eqs.~\ref{eq:Slambda} and \ref{eq:SH2O} are modeled as
\begin{equation}
\label{eq:Sad}
S_\mathrm{ad} =
\begin{dcases*}
\gamma_\mathrm{a}(\lambda_\mathrm{eq}-\lambda)/V_\mathrm{m} & if $\lambda < \lambda_\mathrm{eq}$ (absorption)\\
\gamma_\mathrm{d}(\lambda_\mathrm{eq}-\lambda)/V_\mathrm{m} & if $\lambda > \lambda_\mathrm{eq}$ (desorption)
\end{dcases*}
\end{equation}
where $\lambda_\mathrm{eq}$ is the equilibrium hydration number of the membrane (Sec.~\ref{sec:water_uptake}) and $\gamma_\mathrm{a,d}=k_\mathrm{a,d}/L^\mathrm{CL}$ are the absorption and desorption rates with $L^\mathrm{CL}$ the catalyst layer thickness (Sec.~\ref{sec:compression}) and $k_\mathrm{a,d}$ the ab-/desorption mass transfer coefficients (Sec.~\ref{sec:vapor_sorption}). Similarly, the phase change between vapor and liquid water is modeled as
\begin{equation}
\label{eq:Sec}
S_\mathrm{ec} =
\begin{dcases*}
\gamma_\mathrm{e}(y_{\mathrm{H}_2\mathrm{O}}-y_\mathrm{sat})C & if $y_{\mathrm{H}_2\mathrm{O}} < y_\mathrm{sat}$ (evaporation)\\
\gamma_\mathrm{c}(y_{\mathrm{H}_2\mathrm{O}}-y_\mathrm{sat})C & if $y_{\mathrm{H}_2\mathrm{O}} > y_\mathrm{sat}$ (condensation)
\end{dcases*}
\end{equation}
in which $\gamma_\mathrm{e,c}$ are evaporation/condensation rates (Sec.~\ref{sec:evap_cond}) and $y_\mathrm{sat}=P_\mathrm{sat}/P$ denotes the saturation mole fraction of water vapor with $P_\mathrm{sat}$ the saturation pressure (Sec.~\ref{sec:water_uptake}).

\subsection{Contact resistance}

Electrical contact resistance between individual components of a typical PEMFC is known to have a considerable stake in the overall fuel cell performance loss \cite{cindrella:09}. Here, we follow a recently demonstrated approach \cite{vetter:17} to include both electrical (ECR) and thermal contact resistance (TCR) in the model. The continuity assumptions of the electric phase potential $\varphi_\mathrm{e}$ and temperature $T$ at all affected layer boundaries are replaced by the constraints
\begin{equation}
\label{eq:cr}
\begin{split}
-j_\mathrm{e}\cdot n\bigr\rvert_\mathrm{interface} &= \frac{\varphi_\mathrm{e}^+-\varphi_\mathrm{e}^-}{R_\mathrm{e}}\\
-j_T\cdot n\bigr\rvert_\mathrm{interface} &= \frac{T^+-T^-}{R_T}
\end{split}
\end{equation}
where $[\cdot]^+$ ($[\cdot]^-$) represents the value of the dependent variable $[\cdot]$ on the positive (negative) side of the interface as defined by the unit interface normal vector $n$, whereas $R_\mathrm{e}$ ($R_T$) is the compression-dependent electrical (thermal) contact resistivity of the interface (see Sec.~\ref{sec:compression}). ECR and TCR are modeled at the interior CL/GDL interfaces as well as at the exterior boundaries of the model, at the contact points between the GDLs and the bipolar plates (BPs), the latter of which are not an explicit part of the model. The PEM/CL interfaces are assumed to be perfectly thermally conductive and electrically insulating.

\subsection{Boundary conditions}

At the interfaces between the CLs and the membrane, zero-flux BCs are set for the electron flux as well as at the GDL/CL interfaces for the proton flux. The phase potentials $\varphi_\mathrm{e}$ and $\varphi_\mathrm{p}$ can freely be offset together, because only their derivatives and differences appear in the governing equations. We set $\varphi_\mathrm{e}=0$ at the anode current collector and impose the total cell voltage $\varphi_\mathrm{e}=U$ at the cathode current collector. Through electrical contact resistance (Eq.~\ref{eq:cr}), this amounts to a coupling constraint between the electron flux and electron phase potential at each of the outer GDL surfaces. The temperatures of the anode and cathode bipolar plates, $T_\mathrm{A}$ and $T_\mathrm{C}$, are assumed to be fixed. Water transport in dissolved form is bound to the ionomer phase. Therefore, zero-flux BCs are imposed on $j_\lambda$ at the GDL/CL interfaces. Since gas convection is neglected ($\nabla P\equiv0$), the total gas pressure is given by the gas channel pressures $P_\mathrm{A,C}$. Dirichlet BCs are specified for the mole fractions of the gas species to match these conditions at the GDL/flow channel interfaces: $y_{\mathrm{H}_2} = 1-y_{\mathrm{H}_2\mathrm{O}}$ at the outer AGDL surface and $y_{\mathrm{O}_2} = \alpha_{\mathrm{O}_2}(1-y_{\mathrm{H}_2\mathrm{O}})$ at the outer CGDL surface, where $\alpha_{\mathrm{O}_2}$ denotes the mole fraction of oxygen in the dry supplied oxidation gas. At the same interfaces we additionally specify the relative gas humidity: $y_{\mathrm{H}_2\mathrm{O}} = \mathrm{RH}_\mathrm{A} P_\mathrm{sat}(T_\mathrm{A})/P_\mathrm{A}$ at the outer AGDL surface and $y_{\mathrm{H}_2\mathrm{O}} = \mathrm{RH}_\mathrm{C} P_\mathrm{sat}(T_\mathrm{C})/P_\mathrm{C}$ at the outer CGDL surface. No crossover of gas or liquid water through the membrane is considered, i.e., zero-flux BCs are set at both PEM/CL interfaces for all gas species and the liquid water saturation. At the CGDL/gas channel interface, the liquid water saturation is assumed to coincide with the immobile saturation $s_\mathrm{im}$ (see Sec.~\ref{sec:liquid_water}).

\section{Constitutive parameterization}
\label{sec:parameterization}

\subsection{Electrochemical reaction}
\label{sec:reaction}

The most common approach to model the reaction kinetics is the Butler--Volmer equation
\begin{equation}
\label{eq:butler_volmer}
S_\mathrm{A,C} = S_\mathrm{A,C}^0 \left(\exp\left[\frac{\alpha_\mathrm{A,C}F\eta_\mathrm{A,C}}{RT}\right] - \exp\left[-\frac{\widetilde{\alpha}_\mathrm{A,C}F\eta_\mathrm{A,C}}{RT}\right]\right)
\end{equation}
where $S_\mathrm{A,C}^0=j^0_\mathrm{A,C} a_\mathrm{A,C}(1-s)$ is the product of exchange current density, reactive surface area density and a correction factor ($1-s$) for site blockage by liquid water. $\alpha_\mathrm{A,C}$ ($\widetilde{\alpha}_\mathrm{A,C}$) are the forward (backward) half-reaction transfer coefficients. Measurement data for the hydrogen oxidation reaction suggests that they sum up to unity in the anode, at least for moderate current densities \cite{neyerlin:07}. For the oxygen reduction reaction, on the other hand, there is no consensus in the literature on whether $\alpha_\mathrm{C}=\widetilde{\alpha}_\mathrm{C}=1$ holds, or whether there is a doubling of Tafel slope at intermediate voltages ($\alpha_\mathrm{C}=1$ to $\alpha_\mathrm{C}=0.5$) \cite{neyerlin:06}. We use the former. The exchange current densities can be written as \cite{neyerlin:06,neyerlin:07}
\begin{equation}
\label{eq:j0}
j^0_\mathrm{A,C} = j^{0,\mathrm{ref}}_\mathrm{A,C} \left(\frac{p_\mathrm{A,C}}{P_\mathrm{ref}}\right)^{\delta_\mathrm{A,C}}\exp\left[\frac{E_\mathrm{A,C}}{R}\left(\frac{1}{T_\mathrm{ref}}-\frac{1}{T}\right)\right]
\end{equation}
where $j^{0,\mathrm{ref}}_\mathrm{A,C}$ are the exchange current densities at reference conditions ($P_\mathrm{ref}=1\,\mathrm{atm}$ and $T_\mathrm{ref}=80^\circ\mathrm{C}$), $\delta_\mathrm{A,C}$ the kinetic reaction orders, $E_\mathrm{A,C}$ the half-reaction activation energies, and finally, $p_\mathrm{A}=p_{\mathrm{H}_2}$ and $p_\mathrm{C}=p_{\mathrm{O}_2}$ are the reactant gas partial pressures in the two electrodes. The electrochemical parameters of the model are summarized in Tab.~\ref{tab:electrochemistry}.

\begin{table}
	\centering
	\caption{Electrochemical model parameters.}
	\label{tab:electrochemistry}
	\scalebox{\tabscale}{\begin{tabular}{lrc}
	\toprule
	Parameter & Value & Source\\
	\midrule
	$a_\mathrm{A}$ & $14\,\mathrm{m}_\mathrm{Pt}^2\,\mathrm{cm}^{-3}$ & \cite{flueckiger:09}\\
	$a_\mathrm{C}$ & $28\,\mathrm{m}_\mathrm{Pt}^2\,\mathrm{cm}^{-3}$ & \cite{flueckiger:09}\\
	$E_\mathrm{A}$ & $16\,\mathrm{kJ\,mol}^{-1}$ & \cite{neyerlin:07}\\
	$E_\mathrm{C}$ & $67\,\mathrm{kJ\,mol}^{-1}$ & \cite{neyerlin:06}\\
	$j^{0,\mathrm{ref}}_\mathrm{A}$ & $0.54\,\mathrm{A\,cm}_\mathrm{Pt}^{-2}$ & \cite{neyerlin:07}\\
	$j^{0,\mathrm{ref}}_\mathrm{C}$ & $2.47\!\times\!10^{-8}\,\mathrm{A\,cm}_\mathrm{Pt}^{-2}$ & \cite{neyerlin:06}\\
	$\alpha_\mathrm{A}$, $\widetilde{\alpha}_\mathrm{A}$ & $0.5$ & \cite{neyerlin:07}\\
	$\alpha_\mathrm{C}$, $\widetilde{\alpha}_\mathrm{C}$ & $1$ & \cite{neyerlin:06}\\
	$\delta_\mathrm{A}$ & $0$ & \cite{neyerlin:07}\\
	$\delta_\mathrm{C}$ & $0.54$ & \cite{neyerlin:06}\\
	$\Delta H^\mathrm{ref}$ & $-285.83\,\mathrm{kJ\,mol}^{-1}$ & \cite{chase:98}\\
	$\Delta S_\mathrm{A}^\mathrm{ref}$ & $0.104\,\mathrm{J\,mol}^{-1}\,\mathrm{K}^{-1}$ & \cite{lampinen:93}\\
	$\Delta S_\mathrm{C}^\mathrm{ref}$ & $-163.3\,\mathrm{J\,mol}^{-1}\,\mathrm{K}^{-1}$ & \cite{lampinen:93}\\
	\bottomrule
	\end{tabular}}
\end{table}

\subsection{Electronic and ionic conductivities}
\label{sec:ionic_conductivity}

Constant values are used for the effective electronic conductivities $\sigma_\mathrm{e}^\mathrm{CL}=390\,\mathrm{S\,m}^{-1}$ (for a catalyst layer with ionomer volume fraction $\epsilon_\mathrm{i}=0.3$) \cite{gode:03} and $\sigma_\mathrm{e}^\mathrm{GDL}=450\,\mathrm{S\,m}^{-1}$ (for a SGL 28 AA compressed by 1\,MPa) \cite{schweiss:16}.

\begin{table*}
	\centering
	\caption{Review of protonic conductivities in vapor-equilibrated Nafion membranes. Below the largest $\lambda$ at which $\sigma_\mathrm{p}(\lambda)=0$, the conductivity is set to vanish, which is omitted here for brevity. RT is short for room temperature.}
	\label{tab:sigma_p}
	\scalebox{\tabscale}{\begin{tabular}{lllll}
	\toprule
	Publication & Protonic conductivity [S\,m$^{-1}$] & Temperature & Activation energy [kJ\,mol$^{-1}$] & Membrane\\
	\midrule[\heavyrulewidth]
	Hsu et al., 1980 \cite{hsu:80} & $16(f_\mathrm{w}(\lambda)-0.1)^{1.5}$ & RT & & EW 1050--1500\\\midrule
	Springer et al., 1991 \cite{springer:91} & $0.5139\max\{1,\lambda\}-0.326$ & 30\,$^\circ$C & 10.54 & N117\\\midrule
	Morris \& Sun, 1993 \cite{morris:93} & $12.5(f_\mathrm{w}(\lambda)-0.06)^{1.95}$ & 23--100\,$^\circ$C & & N117\\\midrule
	Sone et al., 1996 \cite{sone:96} & $-0.145+1.57a-4.55a^2+8.86a^3$ & 80\,$^\circ$C & 1.3 & N117\\\midrule
	Eikerling et al., 1998 \cite{eikerling:98} & $0.07+7(f_\mathrm{w}(\lambda)/f_\mathrm{w}(22)-0.1)$ & & & N117\\\midrule
	Thampan et al., 2000 \cite{thampan:00} & $\sigma_0(\lambda,T)(f_\mathrm{w}(\lambda)-f_\mathrm{w}(1.9))^{1.5}$ & & & EW 1100\\\midrule
	Costamagna, 2001 \cite{costamagna:01} & $0.58\lambda-0.5$ & 30\,$^\circ$C & 10.54 (from \cite{springer:91}) & N117\\\midrule
	Edmondson \& Fontanella, 2002 \cite{edmondson:02} & $27.2(f_\mathrm{w}(\lambda)-0.03)^{1.38}$ & RT & & N117\\\midrule
	Kulikovsky, 2003 \cite{kulikovsky:03} & $0.5738\lambda-0.7192$ & 80\,$^\circ$C\\\midrule
	Weber \& Newman, 2004 \cite{weber:04a} & $50(f_\mathrm{w}(\lambda)-0.06)^{1.5}$ & 30\,$^\circ$C & 15 & N117\\\midrule
	Meier \& Eigenberger, 2004 \cite{meier:04} & $0.46\lambda-0.25$ & 25\,$^\circ$C & 9.894 & N117\\\midrule
	Yang et al., 2004 \cite{yang:04} & $1.3\!\times\!10^{-5}\exp[14a^{0.2}]$ & 80--140\,$^\circ$C & & N115\\\midrule
	Choi et al., 2005 \cite{choi:05b} & $\sigma_\Sigma(a,\lambda,T)+\sigma_\mathrm{G}(a,\lambda,T)+\sigma_\mathrm{E}(a,\lambda,T)$\\\midrule
	Fimrite et al., 2005 \cite{fimrite:05} & $\sigma_0(\lambda,T)(f_\mathrm{w}(\lambda)-f_\mathrm{w}(1.65))^{1.5}$ & & & EW 1100\\\midrule
	Hwang et al., 2009 \cite{hwang:09} & $\begin{cases}0.75(\lambda-2.3)&\lambda<5\\0.41(\lambda-5)+3&\lambda\geq5\end{cases}$ & 30\,$^\circ$C & 10.54 (from \cite{springer:91}) & N117\\\midrule
	Maldonado et al., 2012 \cite{maldonado:12} & $-2.91+23.61a-46.09a^2+40.98a^3$ & 80\,$^\circ$C & $13.9a^2-8.87a+11.8$ & N115\\\midrule
	Zhao et al., 2012 \cite{zhao:12} & $77(f_\mathrm{w}(\lambda)-0.1)^2$ & 80\,$^\circ$C & & EW 1100\\
	\bottomrule
	\end{tabular}}
\end{table*}

The protonic conductivity of PFSA membranes and the ionomer phase in the catalyst layers has been the subject of more than 200 papers, with the majority focusing primarily on Nafion with equivalent weight (EW) 1100 \cite{kusoglu:17}. For this reason, it is the first MEA material property that we pay particular attention to here, to examine the uncertainty in PEMFC performance prediction arising from scatter in experimental data on this material property. Being a strong function of the state of ionomer hydration, $\sigma_\mathrm{p}$ dominates the total ohmic losses in the cell in regions where the membrane is relatively dry. In Tab.~\ref{tab:sigma_p}, we summarize published parameterizations for Nafion 1100 EW, which can be (and have been) plugged into numerical models. Where an Arrhenius expression is used to account for temperature dependence, the reported activation energies are also given. Hsu et al.~\cite{hsu:80} originally carried a result over from percolation theory to express the protonic conductivity by a shifted power law, such that one can write
\begin{equation}
\label{eq:sigma_p}
\sigma_\mathrm{p} = M_\mathrm{i}\sigma_0(T)\max\left\{f_\mathrm{w}-f_0,0\right\}^\beta
\end{equation}
where
\begin{equation}
\label{eq:Mi}
M_\mathrm{i}=\frac{\epsilon_\mathrm{i}}{\tau_\mathrm{i}^2}
\end{equation}
is the microstructure factor of the ionomer \cite{holzer:17a} ($\epsilon_\mathrm{i}=\tau_\mathrm{i}=1$ in the PEM and $\epsilon_\mathrm{i}=0.3$, $\tau_\mathrm{i}=1.4$ \cite{babu:16} in the CLs). The water volume fraction in the hydrated ionomer is given by
\begin{equation}
f_\mathrm{w} = \frac{\lambda V_\mathrm{w}}{\lambda V_\mathrm{w}+V_\mathrm{m}}.
\end{equation}
Later models by Thampan et al.~\cite{thampan:00} and Fimrite et al.~\cite{fimrite:05} have extended the percolation-based conductivity by additionally expressing the prefactor $\sigma_0(T)$ as a function of $\lambda$. Springer et al.~\cite{springer:91}, on the other hand, proposed a piecewise linear law in $\lambda$ in their seminal modeling work -- a simple correlation that is still widely used today. The phenomenological parameterizations used by Sone et al.~\cite{sone:96}, Yang et al.~\cite{yang:04} and Maldonado et al.~\cite{maldonado:12} use the water vapor activity (i.e., relative humidity) $a$ to fit the observed ionic conductivity. To implement these into our model, the water activity in the bulk membrane is calculated by inverting the sorption isotherm $\lambda(a)$ (see Sec.~\ref{sec:water_uptake}). Yang's relationship is excluded from our following analysis though, as it predicts far larger conductivities than all others and is considered to be an outlier for this reason.

\begin{figure}
	\centering
	\includegraphics{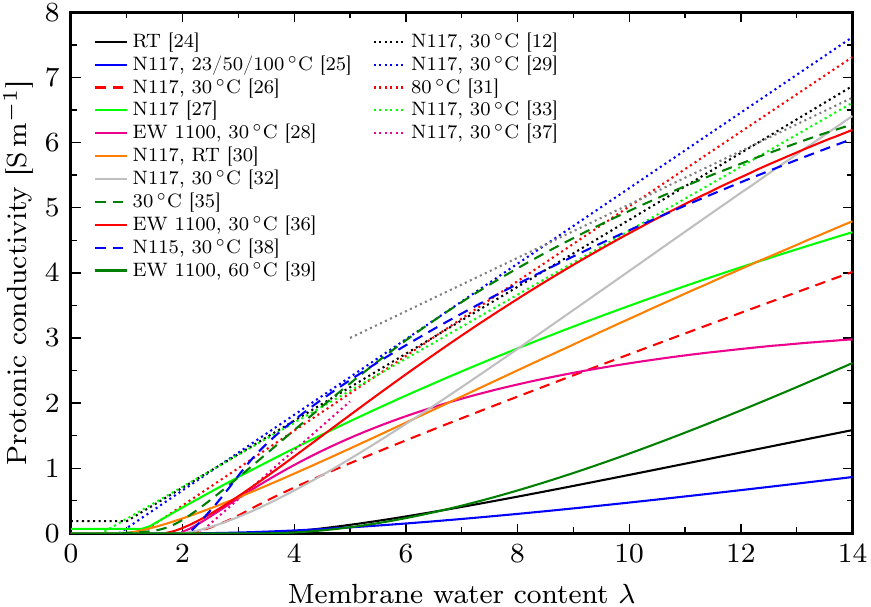}
	\caption{Proposed parameterizations for protonic conductivity as a function of water content in Nafion membranes. Solid lines denote explicit functional relationships $\sigma_\mathrm{p}(\lambda)$, dashed lines are parametric curves $\{\lambda(a),\sigma_\mathrm{p}(a)\}$, dotted lines are purely empirical fits which are piecewise linear in $\lambda$. RT is short for room temperature.}
	\label{fig:sigma_p}
\end{figure}

These proposed parameterizations of $\sigma_\mathrm{p}$ are juxtaposed in Fig.~\ref{fig:sigma_p} to show that they vary considerably, although some spread may certainly partially be the result of different measurement temperatures. Despite characterizing the same class of PFSA membranes, the correlations by Hsu et al.~\cite{hsu:80}, Morris \& Sun \cite{morris:93}, and Zhao et al.~\cite{zhao:12} predict far lower conductivity than the others. As noted by Weber et al.~\cite{weber:04a,kusoglu:17}, the percolation model fits most measured data best. Their coefficients ($f_0=0.06$, $\beta=1.5$, $\sigma_0=50\,\mathrm{S\,m}^{-1}\!\times\!\exp[E_\sigma/R(1/T_\mathrm{ref}-1/T)]$, $E_\sigma=15\,\mathrm{kJ\,mol}^{-1}$, $T_\mathrm{ref}=30^\circ\mathrm{C}$) are used here as the baseline parameterization of ionic conductivity.

\subsection{Thermal conductivity}
\label{sec:thermal_conductivity}

The through-plane thermal conductivity $k$ of the individual layers plays a subordinate role in MEA modeling, as will be shown in Part II. We therefore restrict the discussion on $k$ to a single baseline parameterization for each layer type. In humidified Nafion membranes, it can be approximated by a linear interpolation between the thermal conductivities of water and the dry membrane \cite{khandelwal:06}:
\begin{equation}
k^\mathrm{PEM} = f_\mathrm{w} k_\mathrm{w}+\left(1-f_\mathrm{w}\right) k_0^\mathrm{PEM}
\end{equation}
where
\begin{equation}
k_0^\mathrm{PEM} = \left(0.451-0.286\overline{T}\right)\frac{\mathrm{W}}{\mathrm{m}\,\mathrm{K}}
\end{equation}
is a linear approximation to the thermal conductivity of dry Nafion, and
\begin{multline}
k_\mathrm{w} = \Big(1.6630\overline{T}^{-1.15}-1.7781\overline{T}^{-3.4}+1.1567\overline{T}^{-6.0}\\-0.432115\overline{T}^{-7.6}\Big)\frac{\mathrm{W}}{\mathrm{m}\,\mathrm{K}}
\end{multline}
is the internationally recommended correlation of the thermal conductivity of liquid water at $1\,\mathrm{bar}$ up to $110\,^\circ\mathrm{C}$ with $\overline{T}=T/300\,\mathrm{K}$ \cite{huber:12}. A detailed review of available experimental data on the thermal conductivity of the remaining porous media can be found in \cite{zamel:13}. Here, the Maxwell--Eucken equation \cite{yuan:13}
\begin{equation}
\label{eq:maxwell_eucken}
k = k_\mathrm{s}\frac{2k_\mathrm{s}+k_\mathrm{f}-2(k_\mathrm{s}-k_\mathrm{f})\epsilon_\mathrm{p}}{2k_\mathrm{s}+k_\mathrm{f}+(k_\mathrm{s}-k_\mathrm{f})\epsilon_\mathrm{p}}
\end{equation}
is used, where $k_\mathrm{s}$ is the theoretical conductivity of the solid bulk material, $k_\mathrm{f}$ the conductivity of the fluid filling the pore space, and $\epsilon_\mathrm{p}$ denotes the pore volume fraction. In order to obtain the bulk conductivity $k_\mathrm{s}$ from effective conductivity measurements on real MEA materials, one can invert Eq.~\ref{eq:maxwell_eucken} using $k_\mathrm{f}=0$ or $k_\mathrm{f}\approx0.003\,\mathrm{W\,m}^{-1}\,\mathrm{K}^{-1}$, depending on whether the experiment was conducted in vacuum or air. For vacuum, this yields
\begin{equation}
k_\mathrm{s} = k_0\frac{2+\epsilon_\mathrm{p}}{2(1-\epsilon_\mathrm{p})}
\end{equation}
where $k_0$ is the effectively measured thermal conductivity of the dry porous layer. Alhazmi et al.~\cite{alhazmi:13,alhazmi:14} have conducted the most comprehensive measurements of thermal conductivity of dry GDLs, $k_0^\mathrm{GDL}$, as a function of temperature, clamping pressure and polytetrafluoroethylene content under vacuum conditions. We have fitted the following functional relationship to their data for the SGL 10 series, assuming that the correlations with temperature $T$ and clamping pressure $P_\mathrm{cl}$ are independent:
\begin{equation}
k_0^\mathrm{GDL} = \left(0.776-0.430\overline{T}\right)\left(\frac{P_\mathrm{cl}}{P_\mathrm{ref}}\right)^{0.21}\frac{\mathrm{W}}{\mathrm{m}\,\mathrm{K}}
\end{equation}
where $P_\mathrm{ref}=1\,\mathrm{bar}$. A power law was chosen for the pressure dependence because contact resistivities follow the same relationship (see Sec.~\ref{sec:compression}). For the CLs a constant value is used for the dry thermal conductivity, $k_0^\mathrm{CL}=0.22\,\mathrm{W\,m}^{-1}\,\mathrm{K}^{-1}$ \cite{ahadi:17a}, which lies approximately in the middle of the reported range of values in the literature. Humidity dependence is added through Eqs.~\ref{eq:maxwell_eucken} and \ref{eq:kf} in the CLs just like in the GDLs.

For the fluid conductivity $k_\mathrm{f}$, we assume that liquid water and the gas mixture form transport channels in through-plane direction along which heat is transported in parallel:
\begin{equation}
\label{eq:kf}
k_\mathrm{f} = sk_\mathrm{w} + (1-s)k_\mathrm{g}.
\end{equation}
This choice is motivated by measurements on humidified GDLs \cite{burheim:10a,burheim:11b}, which revealed an overall increase in effective thermal conductivity of as much as 50\% at $s=0.25$ and even more at higher saturations.

The gas phase conductivity $k_\mathrm{g}$ depends on the gas composition. We model it as a linear combination of the conductivities of the individual gas components, with the mole fraction as weights and species conductivities $k_X$ from \cite{rowley:07,green:08}:
\begin{equation}
k_\mathrm{g} = \sum_X y_X k_X,\quad X=\mathrm{H}_2, \mathrm{O}_2, \mathrm{H}_2\mathrm{O}, \mathrm{N}_2.
\end{equation}

\subsection{Water diffusivity in the ionomer}
\label{sec:water_diff}

\begin{figure*}
	\centering
	\includegraphics{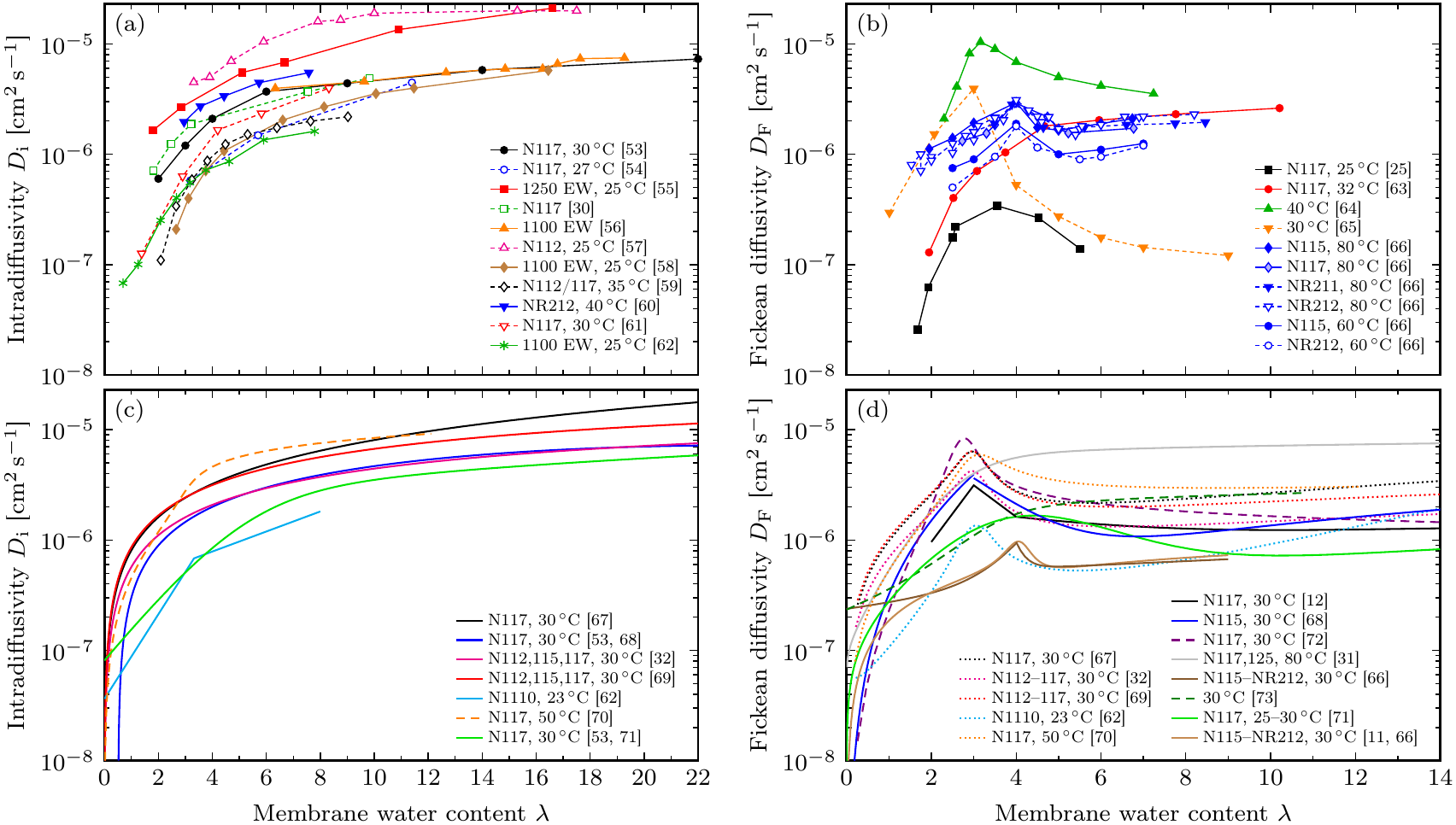}
	\caption{Water diffusivity in Nafion membranes as a function of water content. (a) Experimentally measured intradiffusivity. (b) Experimentally measured Fickean diffusivity. Only direct measurements that don't rely on Eq.~\ref{eq:darken} are shown. (c) Parametric expressions for intradiffusivity. (d) Parametric expressions for Fickean diffusivity. In (c) and (d), solid lines represent explicit functions $D(\lambda)$, dashed lines are parametric curves $\{\lambda(a),D(a)\}$ as a function of activity $a$ and dotted lines are converted from intradiffusivity using the Darken factor (from Springer's isotherm \cite{springer:91} where none is given in the respective publication). All ordinates reach up to the self-diffusivity of water at 25\,$^\circ$C, $2.3\!\times\!10^{-5}\,\mathrm{cm}^2\,\mathrm{s}^{-1}$ \cite{mills:71,holz:00}.}
	\label{fig:D_lambda}
\end{figure*}

\begin{table*}
	\centering
	\caption{Review of water diffusion coefficients in vapor-equilibrated Nafion membranes.}
	\label{tab:D_lambda}
	\scalebox{\tabscale}{\begin{tabular}{lllll}
	\toprule
	Publication & Water diffusivity $D_\mathrm{i}$ or $D_\mathrm{F}$ [$10^{-6}$\,cm$^2$\,s$^{-1}$] & Temperature & Activation energy & Membrane\\
	\midrule[\heavyrulewidth]
	Springer et al., 1991 \cite{springer:91} & $D_\mathrm{F}=2.563-0.33\lambda+0.0264\lambda^2-0.000671\lambda^3$ for $\lambda>4$ & 30\,$^\circ$C & 20.1\,kJ\,mol$^{-1}$ & N117\\\midrule
	Fuller, 1992 \cite{fuller:92b} & $D_\mathrm{i}=3.5\!\times\!10^4\lambda/14$ & $\infty$ & 20.3\,kJ\,mol$^{-1}$ & N117\\\midrule
	Motupally et al., 2000 \cite{motupally:00,zawodzinski:91} & $D_\mathrm{i}=0.631(-0.501+\lambda-0.0209\lambda^2)$ & 30\,$^\circ$C & & N117\\
	& $D_\mathrm{F}=\begin{cases}3100\lambda(\exp[0.28\lambda]-1) & \lambda<3\\417\lambda(161\exp[-\lambda]+1) & \lambda\geq3\end{cases}$ & $\infty$ & 20.3\,kJ\,mol$^{-1}$ & N115\\\midrule
	Ye \& LeVan, 2003 \cite{ye:03} & $D_\mathrm{F}=6.76\widetilde{p}_\mathrm{v}^{1.5}/(5.9673-8.9472\widetilde{p}_\mathrm{v}+4.0622\widetilde{p}_\mathrm{v}^2)$, $\widetilde{p}_\mathrm{v}=p_\mathrm{v}/\mathrm{kPa}$ & 23.5--25\,$^\circ$C\\\midrule
	Kulikovsky, 2003 \cite{kulikovsky:03} & $D_\mathrm{F}=4.1(\lambda/25)^{0.15}(1+\tanh[(\lambda-2.5)/1.4])$ & 80\,$^\circ$C & & N117,125\\\midrule
	Weber \& Newman, 2004 \cite{weber:04a} & $D_\mathrm{i}=18f_\mathrm{w}$ & 30\,$^\circ$C & 20\,kJ\,mol$^{-1}$ & N112,115,117\\\midrule
	Ge et al., 2005 \cite{ge:05} & $D_\mathrm{i}=27.2f_\mathrm{w}$ & 30\,$^\circ$C & 20.1\,kJ\,mol$^{-1}$ & N112,115,117\\\midrule
	Zhao et al., 2011 \cite{zhao:11} & $D_\mathrm{F}$: piecewise exponential in $\lambda$, interpolation in $T$ & 23--70\,$^\circ$C &  & N1110\\\midrule
	Myles et al., 2011 \cite{myles:11} & $D_\mathrm{i}=6.0667a+3.1333a^2$ & 50\,$^\circ$C & & N117\\
	& $D_\mathrm{i}=7.2167a+3.4833a^2$ & 60\,$^\circ$C & & N117\\\midrule
	Mittelsteadt \& Staser, 2011 \cite{mittelsteadt:11} & $D_\mathrm{F}=\begin{cases}732\exp[0.12\lambda]+5.41\exp[1.44\lambda]&\lambda<4\\1.58\!\times\!10^{11}\exp[-4.66\lambda]+1450\exp[0.04\lambda]&\lambda\geq4\end{cases}$ & $\infty$ & 20.3\,kJ\,mol$^{-1}$ & \parbox{5em}{N115,117\\NR211,212}\\\midrule
	Caulk et al., 2012 \cite{caulk:12} & $D_\mathrm{F}=0.032\,\mathrm{mol\,cm}^{-3}\,\mathrm{bar}^{-1}\!\times\!\exp[3.4a]P_\mathrm{sat}V_\mathrm{m}(\partial a/\partial\lambda)$ & 90\,$^\circ$C & 22\,kJ\,mol$^{-1}$\\\midrule
	Lokkiluoto \& Gasik, 2013 \cite{lokkiluoto:13} & Phenomenological expressions for $D_\mathrm{F}(\lambda)$ and $D_\mathrm{i}(\lambda)$ & 30\,$^\circ$C & & N117\\\midrule
	Vetter \& Schumacher, 2018 \cite{vetter:18,mittelsteadt:11} & $\displaystyle{D_\mathrm{F}=\frac{3.842\lambda^3-32.03\lambda^2+67.74\lambda}{\lambda^3-2.115\lambda^2-33.013\lambda+103.37}}$ & 80\,$^\circ$C & 20\,kJ\,mol$^{-1}$ & \parbox{5em}{N115,117\\NR211,212}\\
	\bottomrule
	\end{tabular}}
\end{table*}

Water diffusion within and through the membrane has been a topic of extensive research over the past decades and has been thoroughly reviewed in the works of Kusoglu \& Weber \cite{kusoglu:12b,kusoglu:17}. The numerous studies carried out to measure and parameterize the water diffusivity vary in the transport mode considered (driven by a pressure gradient, concentration gradient, temperature gradient, or chemical potential gradient), in the experimental technique (NMR, QENS, conductivity), in the types of diffusion coefficients being measured, in the studied regime (Fickean vs.\ non-Fickean, steady-state vs.\ transient), and even in the data separation and interpretation (e.g., correcting for measurement device resistance and interfacial resistance or not). As a result, the reported diffusion coefficients of dissolved water in Nafion are scattered over one to two orders of magnitude, which calls for a quantitative analysis of the modeling uncertainty associated with this material property.

As shown in Fig.~\ref{fig:D_lambda}a,c, the intradiffusivity $D_\mathrm{i}$ is a monotonically increasing function of $\lambda$ \cite{kusoglu:17}, whereas most studies agree that the Fickean diffusivity $D_\mathrm{F}$ exhibits a pronounced peak around $\lambda\approx2-4$ (Fig.~\ref{fig:D_lambda}b,d). This local maximum in $D_\mathrm{F}$ stems from the \emph{Darken factor}, which relates the two diffusivities \cite{springer:91}:
\begin{equation}
\label{eq:darken}
D_\mathrm{F} = \frac{\partial\ln a}{\partial\ln\lambda}D_\mathrm{i} = \frac{\lambda}{a}\left(\frac{\partial\lambda}{\partial a}\right)^{-1}D_\mathrm{i}
\end{equation}
where $\lambda(a)$ is the vapor uptake isotherm (see Sec.~\ref{sec:water_uptake}). Since $\lambda(a)$ is relatively flat at $\lambda\approx2-4$, the Fickean diffusivity $D_\mathrm{F}$ peaks there. However, the existence of a local maximum in $D_\mathrm{F}$ is still a subject of controversy \cite{rivin:01,caulk:12,olesen:12}.

$D_\mathrm{F}$ can either be measured directly or calculated from $D_\mathrm{i}$ using Eq.~\ref{eq:darken}. A compilation of published correlations between either of them and the level of hydration (via $a$, $\lambda$ or $f_\mathrm{w}$) is given in Tab.~\ref{tab:D_lambda} and Fig.~\ref{fig:D_lambda}c,d. Given $D_\mathrm{F}$, the effective diffusivity required for Eq.~\ref{eq:j_lambda} can be calculated as
\begin{equation}
D_\lambda=M_\mathrm{i}D_\mathrm{F}
\end{equation}
with $M_\mathrm{i}$ from Eq.~\ref{eq:Mi}. We have recently identified the parametric expression given in the last row of Tab.~\ref{tab:D_lambda} as the most plausible and convenient for numerical modeling of water transport within Nafion membranes \cite{vetter:18}. It is used as the baseline parameterization also for the present study.

\begin{figure}
	\centering
	\includegraphics{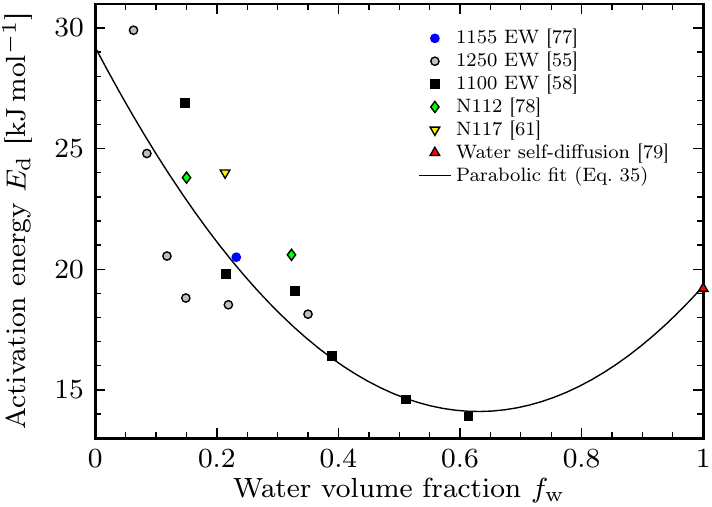}
	\caption{Activation energy of the water diffusion coefficient in Nafion membranes as a function of membrane hydration.}
	\label{fig:D_lambda_energy}
\end{figure}

For the temperature dependence, the Arrhenius equation is assumed to hold almost unanimously in experimental and modeling studies, even though a Speedy--Angell power law fits experimental data better \cite{holz:00}. The research group of Eisenberg \cite{yeo:77,takamatsu:79} was the first to estimate the activation energy $E_\mathrm{d}\approx20\,\mathrm{kJ\,mol}^{-1}$ for a Nafion membrane immersed in liquid water, which is close to the value for water self-diffusion ($19.2\,\mathrm{kJ\,mol}^{-1}$ \cite{wang:51}). This value has subsequently been used by almost all researchers to fit their experimental data and to model $D_\mathrm{F}(\lambda,T)$, neglecting that it might depend on the mode and level of membrane hydration. Later measurements have brought up a variety of other values: 12--16 \cite{kidena:10}, 22 \cite{caulk:12}, 24 \cite{ma:11} and 28\,kJ\,mol$^{-1}$ \cite{zhao:11}. Gong et al.~\cite{gong:01}, Kreuer et al.~\cite{kreuer:08} and Guillermo et al.~\cite{guillermo:09} investigated the dependency of $E_\mathrm{d}$ on hydration. Their findings are juxtaposed in Fig.~\ref{fig:D_lambda_energy}. A polynomial least-squares fit to the available data yields
\begin{equation}
\label{eq:Ed}
E_\mathrm{d} = \left(38.0f_\mathrm{w}^2-47.9f_\mathrm{w}+29.2\right)\,\mathrm{kJ\,mol}^{-1},
\end{equation}
which is the activation energy implemented in the present model to adapt the rational polynomial in $\lambda$ from Tab.~\ref{tab:D_lambda} to arbitrary temperatures in the baseline parameterization.

For a Nafion membrane of type NR211, which is considered for the baseline simulation, $\rho_\mathrm{m}=1.97\,\mathrm{g\,cm}^{-3}$ \cite{nafion:16} and $m_\mathrm{m}=1020\,\mathrm{g\,mol}^{-1}$ \cite{peron:10}, such that the equivalent volume is given by $V_\mathrm{m}=m_\mathrm{m}/\rho_\mathrm{m}\approx517.8\,\mathrm{cm}^3\,\mathrm{mol}^{-1}$.

\subsection{Electro-osmosis}
\label{sec:electro_osmosis}

\begin{table*}
	\centering
	\caption{Review of electro-osmotic drag coefficients in vapor-equilibrated Nafion membranes.}
	\label{tab:xi}
	\scalebox{\tabscale}{\begin{tabular}{lllll}
	\toprule
	Publication & Electro-osmotic drag coefficient $\xi$ & Temperature & Membrane\\
	\midrule[\heavyrulewidth]
	Springer et al., 1991 \cite{springer:91} & $2.5\lambda/22$ & 30\,$^\circ$C & N117\\\midrule
	Fuller \& Newman, 1992 \cite{fuller:92} & $-BCa\exp[-Ca]$, $B=-3.7206$, $C=1.339$ & 25--37.5\,$^\circ$C & N117\\\midrule
	Fuller, 1992 \cite{fuller:92b} & $((0.35\lambda)^{-4}+1.47^{-4})^{-1/4}$ & 25--37.5\,$^\circ$C & N117\\\midrule
	Eikerling et al., 1998 \cite{eikerling:98} & $1.2+1.3(f_\mathrm{w}(\lambda)/f_\mathrm{w}(22))^5$ & & N117\\\midrule
	Dutta et al., 2001 \cite{dutta:01} & $0.0029\lambda^2+0.05\lambda$\\\midrule
	Kulikovsky, 2003 \cite{kulikovsky:03,vanbussel:98} & $\max\{1,0.117\lambda-0.0544\}$ & 80\,$^\circ$C\\\midrule
	Weber \& Newman, 2004 \cite{weber:04a} & $\min\{1,\lambda\}$ & 30\,$^\circ$C\\\midrule
	Meier \& Eigenberger, 2004 \cite{meier:04} & $1+0.028\lambda+0.0026\lambda^2$ & 25\,$^\circ$C\\\midrule
	Ge et al., 2006 \cite{ge:06} & polynomial in $\lambda$, linear interpolation in $T$ & 30--80\,$^\circ$C & N117\\\midrule
	Lokkiluoto \& Gasik, 2013 \cite{lokkiluoto:13} & $\sqrt{\lambda}/2$ & & N117\\
	\bottomrule
	\end{tabular}}
\end{table*}

\begin{figure}
	\centering
	\includegraphics{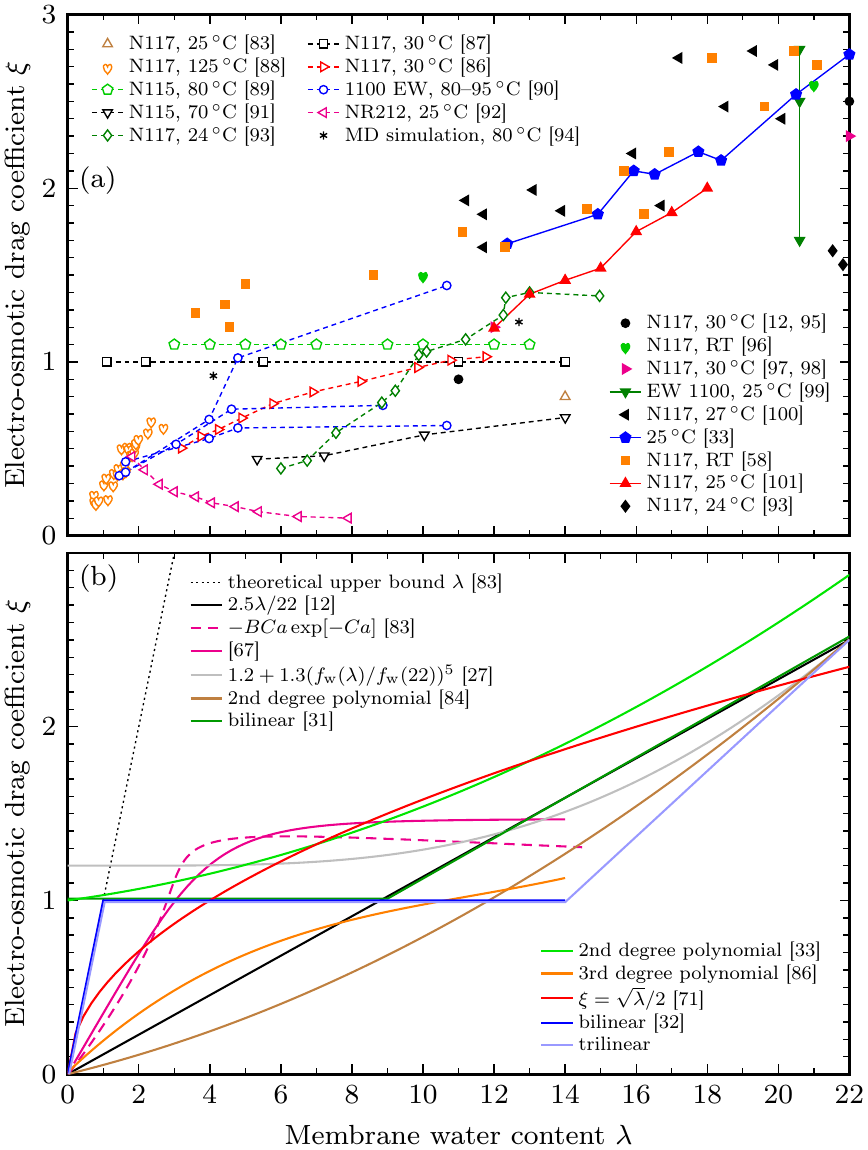}
	\caption{Electro-osmotic drag coefficient in Nafion membranes as a function of water content. (a) Experimental data. Open symbols denote measurements on vapor-equilibrated, closed symbols on liquid-equilibrated (and subsequently dried) membranes. (b) Proposed parameterizations. Solid lines denote explicit functional relationships $\xi(\lambda)$, the dashed line is a parametric curve $\{\lambda(a),\xi(a)\}$.}
	\label{fig:xi}
\end{figure}

Electro-osmotic drag in Nafion has been another subject of controversial debate owing to the complexity of coupled ion/water transport and the difficulty in measuring it \cite{dai:09,kusoglu:17}. Springer et al.~\cite{springer:91} proposed a linear approximation for the electro-osmotic drag coefficient
\begin{equation}
\label{eq:xi}
\xi=\xi_\mathrm{l}\lambda/\lambda_\mathrm{l}
\end{equation}
based on a single data point for Nafion 117 immersed in liquid water: ($\xi_\mathrm{l},\lambda_\mathrm{l})=(2.5,22)$. This relationship is still widely used in MEA modeling. Later measurements \cite{zawodzinski:95} have led researchers to conclude that vapor-equilibrated PFSA membranes are more appropriately characterized by $\xi\equiv 1$.

Fig.~\ref{fig:xi} and Tab.~\ref{tab:xi} provide an overview over published measurements and parameterizations of the electro-osmotic drag coefficient, from which it becomes clear that a conclusive reliable correlation $\xi(\lambda,T)$ is still missing. Care must be taken when interpreting Fig.~\ref{fig:xi}a, because some of these measurements were performed on liquid-equilibrated membranes with different degrees of drying to obtain mid-range water contents. These data might not be representative for vapor sorption. Agreement has not even been found on the general trend of the drag coefficient, with data suggesting an increasing, invariant, or even decreasing value of $\xi$ with increasing membrane hydration.

A frequently used class of parameterizations are piecewise linear functions \cite{kulikovsky:03,bussel:98,meng:04}, which take the form
\begin{equation}
\label{eq:xi_multilinear}
\xi = 
\begin{dcases}
\lambda & 0 \leq \lambda < 1\\
1 & 1\leq\lambda\leq\lambda_\mathrm{v}(1)\\
1+(\xi_\mathrm{l}-1)\frac{\lambda-\lambda_\mathrm{v}(1)}{\lambda_\mathrm{l}-\lambda_\mathrm{v}(1)} & \lambda_\mathrm{v}(1)<\lambda\leq\lambda_\mathrm{l}
\end{dcases}
\end{equation}
when taking the theoretical upper bound $\xi\leq\lambda$ \cite{fuller:92} into account. Here, $\lambda_\mathrm{l}$ is the water content for a liquid-equilibrated membrane, $\lambda_\mathrm{v}(1)$ the water content for a vapor-equilibrated membrane at unit vapor activity (see Sec.~\ref{sec:water_uptake}), and $\xi_\mathrm{l}$ the drag coefficient for a liquid-equilibrated membrane. Eq.~\ref{eq:xi_multilinear} is a simple attempt at taking Schroeder's paradox ($\lambda_\mathrm{l}\gg\lambda_\mathrm{v}(1)$, see Sec.~\ref{sec:water_uptake}) into account in a global parameterization $\xi(\lambda)$. Weber \& Newman \cite{weber:04a} suggested to use the Arrhenius equation
\begin{equation}
\label{eq:xi_l}
\xi_\mathrm{l} = 2.55\exp\left[\frac{E_\xi}{R}\left(\frac{1}{T_\mathrm{ref}}-\frac{1}{T}\right)\right]
\end{equation}
with $T_\mathrm{ref}=30^\circ\mathrm{C}$ and an activation energy of $E_\xi=4\,\mathrm{kJ\,mol}^{-1}$ to model the temperature dependence of $\xi_\mathrm{l}$.

In summary, Springer's linear law (Eq.~\ref{eq:xi}) matches the widely scattered experimental data best. It is therefore is used here as the baseline parameterization, together with Eq.~\ref{eq:xi_l} for temperature dependence.

\subsection{Thermo-osmosis}
\label{sec:thermo_osmosis}

Thermo-osmotic transport of dissolved water is an entropic effect and occurs from cold to hot regions in hydrophilic membranes, i.e., in direction of the positive temperature gradient \cite{tasaka:90,barragan:17}. The thermo-osmotic transport coefficient $D_T$ is thus negative for typical PFSA membranes \cite{villaluenga:06,kim:09a}. Kim \& Mench \cite{kim:09a} studied the temperature dependence of thermo-osmosis in liquid-equilibrated Nafion 112, Flemion SH50 and Gore-Select membranes and reported that the Arrhenius equation holds with an activation energy that is indistinguishable from that of concentration gradient-driven diffusion. This suggests that the transport mechanism for thermo-osmosis might be the same as for diffusion, albeit with different driving force. We therefore fitted their measured values of the thermo-osmotic transport coefficient in N112 using $T_\mathrm{ref}=80^\circ\mathrm{C}$ and the activation energy $E_\mathrm{d}$ from Eq.~\ref{eq:Ed} and obtained
\begin{equation}
D_T(T) = -7.2\!\times\!10^{-7}\,\mathrm{mol}\,\mathrm{m}^{-1}\,\mathrm{s}^{-1}\,\mathrm{K}^{-1}\,\exp\left[\frac{E_\mathrm{d}}{R}\left(\frac{1}{T_\mathrm{ref}}-\frac{1}{T}\right)\right].
\end{equation}

Measurements of $D_T$ for \textit{vapor-equilibrated} PFSA membranes are still missing in the literature. The magnitude of $D_T$ in various anion exchange membranes has been reported to increase with growing water content \cite{tasaka:92,suzuki:94}, and if the transport mechanism of thermo-osmosis is indeed similar to diffusion, it appears natural to assume that $D_T\to0$ as $\lambda\to0$. Therefore, in order not to overestimate the effect of thermo-osmosis much in the model for vapor-equilibrated membranes, a linear approximation between zero and the reported coefficient for liquid-equilibrated Nafion is used here, analogous to Springer's linear interpolation of the electro-osmotic drag coefficient in Eq.~\ref{eq:xi}:
\begin{equation}
\label{eq:DT}
D_T(\lambda,T) = \frac{\lambda}{\lambda_\mathrm{l}} D_T(T).
\end{equation}

\subsection{Membrane hydration}
\label{sec:water_uptake}

\begin{table*}
	\centering
	\caption{Review of water vapor sorption isotherms of Nafion membranes. Only parameterizations which apply to the full vapor activity range $a\in[0,1]$ are included.}
	\label{tab:lambda_v}
	\scalebox{\tabscale}{\begin{tabular}{llll}
	\toprule
	Publication & Equilibrium water content $\lambda_\mathrm{v}$ & Temperature & Membrane\\
	\midrule[\heavyrulewidth]
	Springer et al., 1991 \cite{springer:91} & $0.043+17.81a-39.85a^2+36.0a^3$ & 30\,$^\circ$C & N117\\\midrule
	Hinatsu et al., 1994 \cite{hinatsu:94} & $0.300+10.8a-16.0a^2+14.1a^3$ & 80\,$^\circ$C & N117,125\\\midrule
	Futerko \& Hsing, 1999 \cite{futerko:99} & numerical solution of the implicit equation\\
	& $(1-f_\mathrm{mb})\exp[(1-1/r)f_\mathrm{mb}+\chi f_\mathrm{mb}^2]=a$\\
	& with $f_\mathrm{mb}=(r+\lambda_\mathrm{b})/(r+\lambda_\mathrm{v})$, $\lambda_\mathrm{b}=Ka/(1+Ka)$, $r=V_\mathrm{m}/V_\mathrm{w}$,\\
	& $\chi=1.936-2.18\,\mathrm{kJ\,mol}^{-1}/RT$, $K=0.0256\exp[22.4\,\mathrm{kJ\,mol}^{-1}/RT]$ & & N117\\\midrule
	Thampan et al., 2000 \cite{thampan:00} & $\lambda_\mathrm{v}=\lambda_\mathrm{BET}=\lambda_\mathrm{m}Ka/(1-a)\!\times\!(1-a^n-na^n(1-a))/(Ka(1-a^n)+1-a)$\\
	& with $\lambda_\mathrm{m}=1.8$, $K=150$, $n=13.5$ & 25--30\,$^\circ$C & N117\\\midrule
	Meyers \& Newman, 2002 \cite{meyers:02} & $\lambda_\mathrm{v}=\lambda_2(1+\exp[0.3-\lambda_2])$ where $\lambda_2$ solves the coupled implicit equations\\
	and & $\begin{cases}\lambda_3\exp[f_1\lambda_3+f_2\lambda_2]=K_1(1-\lambda_3)(\lambda_2-\lambda_3)\\K_2(\lambda_2-\lambda_3)\exp[f_2\lambda_3+f_3\lambda_2]=a\end{cases}$\\
	Weber \& Newman, 2004 \cite{weber:04a} & with $K_1=100$, $K_2=0.217\exp[1\,\mathrm{kJ\,mol}^{-1}/R\!\times\!(1/303.15\,\mathrm{K}-1/T)]$,\\
	& $f_1 = 2(m_{22}-2m_{31}-2m_{23})/m_\mathrm{m}$, $f_2 = 2(m_{23}-m_{22})/m_\mathrm{m}$, $f_3 = 2m_{22}/m_\mathrm{m}$, & & N117\\
	& $m_{22} = -41.7\,\mathrm{g\,mol}^{-1}$, $m_{23} = -52.0\,\mathrm{g\,mol}^{-1}$, $m_{31} = -3721.6\,\mathrm{g\,mol}^{-1}$\\\midrule
	Kulikovsky, 2003 \cite{kulikovsky:03} & $0.3+6a(1-\tanh[a-0.5])+3.9\sqrt{a}(1+\tanh[(a-0.89)/0.23])$ & 80\,$^\circ$C & N117,125\\\midrule
	Choi \& Datta, 2003 \cite{choi:03} & numerical solution of the implicit equation\\
	& $(\lambda_\mathrm{v}-\lambda_\mathrm{b})/(1+\lambda_\mathrm{v}-\lambda_\mathrm{b})=a\exp[-V_\mathrm{w}P/RT]$ with $\lambda_\mathrm{b}=\lambda_\mathrm{BET}$, $\lambda_\mathrm{m}=1.8$,\\
	& $K=100$, $n=5$, $P=\kappa f_\mathrm{w}-a_\mathrm{p}\sigma\cos\theta/f_\mathrm{w}$, $a_\mathrm{p}=210\,\mathrm{m}^2\,\mathrm{cm}^{-3}$, $\kappa=183\,\mathrm{atm}$, & & N117\\
	& $\sigma=72.1\,\mathrm{mN\,m}^{-1}$, $\theta=(116-7.15a+28.4a^2-39.3a^3)\,^\circ$ (fit to data from \cite{zawodzinski:93b})\\\midrule
	Meier \& Eigenberger, 2004 \cite{meier:04} & $17.81a-39.85a^2+35a^3$\\\midrule
	Choi et al., 2005 \cite{choi:05a} & numerical solution of the implicit equation\\
	& $(1-f_\mathrm{mb})\exp[(1-1/r)f_\mathrm{mb}+\chi f_\mathrm{mb}^2]=a\exp[-V_\mathrm{w}P/RT]$ with\\
	& $f_\mathrm{mb}=(r+\lambda_\mathrm{b})/(r+\lambda_\mathrm{v})$, $\lambda_\mathrm{b}=\lambda_\mathrm{BET}$, $\lambda_\mathrm{m}=1.8$, $K=1000$, $n=5$,\\
	& $P=2E(f_\mathrm{m}^{1/3}-f_\mathrm{m}^{7/3})/9-a_\mathrm{p}\sigma\cos\theta/f_\mathrm{w}$, $r=V_\mathrm{m}/V_\mathrm{w}$, $f_\mathrm{m}=1-f_\mathrm{w}$, & & EW 1100\\
	& $a_\mathrm{p}=210\,\mathrm{m}^2\,\mathrm{cm}^{-3}$, $\sigma=72.1\,\mathrm{mN\,m}^{-1}$, $\theta=98\,^\circ$ \\\midrule
	Takata et al., 2007 \cite{takata:07} & $V_\mathrm{m}/m_\mathrm{w}\!\times\!A_\mathrm{L}B_\mathrm{L}p_\mathrm{v}/(1+A_\mathrm{L}p_\mathrm{v})\!\times\!(1+(n-1)(A_\mathrm{C}p_\mathrm{v})^{n-1})$ with\\
	& $\substack{\displaystyle{A_\mathrm{L}=1.53\!\times\!10^{-10}\,\mathrm{Pa}^{-1}\!\times\!\exp[39\,\mathrm{kJ\,mol}^{-1}/RT],}\\\displaystyle{A_\mathrm{C}=2.40\!\times\!10^{-12}\,\mathrm{Pa}^{-1}\!\times\!\exp[46\,\mathrm{kJ\,mol}^{-1}/RT],}}$ $B_\mathrm{L}=0.160\,\mathrm{g\,cm}^{-3}$, $n=5.15$ & 10--80\,$^\circ$C & N117\\\midrule
	Costamagna et al., 2008 \cite{costamagna:08} & $\lambda_\mathrm{m}Kka/(1-ka)/(1+(K-1)ka)$ with $\lambda_\mathrm{m}=2.3$, $K=70$, $k=0.7$ & 20\,$^\circ$C & N117\\\midrule
	Ochi et al., 2009 \cite{ochi:09} & $0.8486+24.594a-112.7a^2+300a^3-358.78a^4+162.77a^5$ & 30--90\,$^\circ$C & N117\\\midrule
	Kusoglu et al., 2009 \cite{kusoglu:09} & same as Choi \& Datta, but with\\
	& $P=E(1-[1+\kappa(f_\mathrm{m}^{-1/3}-1)][1-f_\mathrm{c}^{1/2}]/[1-f_\mathrm{c,dry}^{1/2}])$, $f_\mathrm{c}=f_\mathrm{w}+f_\mathrm{m}f_\mathrm{c,dry}$,\\
	& $E/\mathrm{MPa}=(1000-T/0.4\,\mathrm{K})/(1-f_\mathrm{c,dry}^{1/2})$, $f_\mathrm{m}=1-f_\mathrm{w}$, $f_\mathrm{c,dry}=V_{\mathrm{SO}3}/V_\mathrm{m}$, & 25--85\,$^\circ$C & EW 1100\\
	& $V_{\mathrm{SO}3}=40.94\,\mathrm{cm}^3\,\mathrm{mol}^{-1}$, $\kappa=5.6$\\\midrule
	Mittelsteadt \& Liu, 2010 \cite{mittelsteadt:10} & $(1+0.2325a^2(T/^\circ\mathrm{C}-30)/30)(13.41a-18.92a^2+14.22a^3)$\\\midrule
	Myles et al., 2011 \cite{myles:11} & $16.0674a-32.3781a^2+28.4170a^3$ & 50\,$^\circ$C & N117\\
	& $15.0395a-28.3372a^2+24.4519a^3$ & 60\,$^\circ$C & N117\\\midrule
	Eikerling \& Berg, 2011 \cite{eikerling:11} & $3.0a^{0.2}+11.0a^4$ & 30\,$^\circ$C & N117\\\midrule
	Li et al., 2013 \cite{li:13} & $\lambda_\mathrm{v}=\lambda_\mathrm{DM}=\lambda_\mathrm{m}ka/(1-ka)+\lambda_\mathrm{m}(K-1)ka/(1+(K-1)ka)$ with\\
	& $\lambda_\mathrm{m}=3.1$, $K=11.4$, $k=0.80$ & 25\,$^\circ$C & N117\\
	& $\lambda_\mathrm{m}=3.3$, $K=\phantom{0}3.8$, $k=0.79$ & 20\,$^\circ$C & N117\\
	& $\lambda_\mathrm{m}=3.1$, $K=\phantom{0}9.5$, $k=0.75$ & 50\,$^\circ$C & N112\\\midrule
	Kreuer, 2013 \cite{kreuer:13,kreuer:18} & numerical solution of the implicit equation\\
	& $\lambda_\mathrm{v}=\lambda_\mathrm{f}+\sum_{i=1}^5\prod_{j=1}^i\theta_j$ with $\lambda_\mathrm{f}=1/(PV_\mathrm{w}/RT-\ln a)$, & & N117\\
	& $\theta_j=\exp[\Delta H_j/RT-1/\lambda_\mathrm{f}]/(1+\exp[\Delta H_j/RT-1/\lambda_\mathrm{f}])$, $P=E'(\lambda_\mathrm{v}V_\mathrm{w}/V_\mathrm{m})^{1/3}$\\\midrule
	Didierjean et al., 2015 \cite{didierjean:15} & $0.165+13.86a-24.51a^2+23.01a^3$ & 25\,$^\circ$C & N117\\\midrule
	Shi et al., 2016 \cite{shi:16} & $\lambda_\mathrm{v}=\lambda_\mathrm{DM}$ with $\lambda_\mathrm{m}=2.671$, $K=7.269$, $k=0.7677$ & 25\,$^\circ$C & NR212\\\midrule
	Morin et al., 2017 \cite{morin:17} & $0.053056+41.1263a-180.83a^2+406.89a^3-381.59a^4+69.385a^5+62.335a^6$ & 25\,$^\circ$C & N117\\
	\bottomrule
	\end{tabular}}
\end{table*}

\begin{figure}
	\centering
	\includegraphics{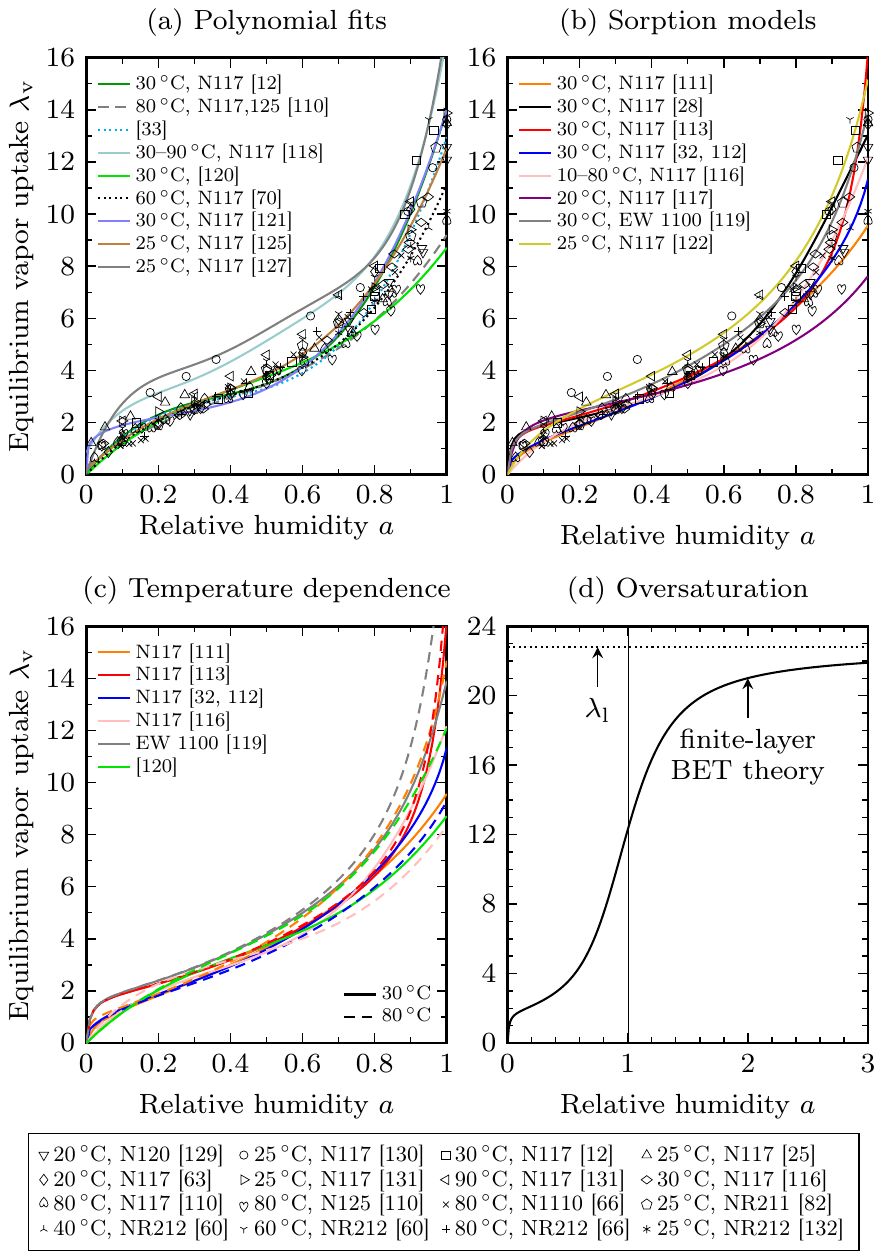}
	\caption{Vapor sorption isotherms for Nafion membranes. Symbols denote data from experimental measurements, lines represent fitted curves at $T=30\,^\circ\mathrm{C}$ (solid) and $T=80\,^\circ\mathrm{C}$ (dashed).}
	\label{fig:lambda_v}
\end{figure}

In the parameterization of the equilibrium water content of the ionomer in Eq.~\ref{eq:Sad}, we account for simultaneous partial contact of the ionomer with liquid water and water vapor as well as for Schroeder's paradox by writing
\begin{equation}
\label{eq:lambda_eq}
\lambda_\mathrm{eq} = s\lambda_\mathrm{l} + (1-s)\lambda_\mathrm{v}
\end{equation}
where $\lambda_\mathrm{l}$ ($\lambda_\mathrm{v}$) denotes the hydration number when the membrane is liquid-equilibrated (vapor-equilibrated). The vapor sorption isotherm $\lambda_\mathrm{v}(a)$ has been the subject of a vast number of experimental and theoretical studies. A chronological listing of proposed explicit and implicit functional relationships is given in Tab.~\ref{tab:lambda_v}. Historically, polynomial fits to experimental data (of degree 3 or higher in $a$) have been popular in PEMFC modeling, starting with Springer's and Hinatsu's measurements at 30 and 80$\,^\circ\mathrm{C}$, respectively. Temperature dependence can be introduced by linearly interpolating between these two polynomials. In Fig.~\ref{fig:lambda_v}a, the different polynomials are plotted together with a selection of experimental data points from the open literature, showing that there is an uncertainty band of around 2 in width over the entire activity range, with a tendency to widen toward saturation. The water vapor activity is calculated as
\begin{equation}
a = \frac{y_{\mathrm{H}_2\mathrm{O}}}{y_\mathrm{sat}} = \frac{p_{\mathrm{H}_2\mathrm{O}}}{P_\mathrm{sat}}
\end{equation}
using \cite{wagner:93}
\begin{multline}
\label{eq:Psat}
P_\mathrm{sat} = P_\mathrm{c}\exp\Bigl[\frac{T_\mathrm{c}}{T}\Bigl(-7.8595\widehat{T}+1.8441\widehat{T}^{1.5}-11.787\widehat{T}^3\\+22.681\widehat{T}^{3.5}-15.962\widehat{T}^4+1.8012\widehat{T}^{7.5}\Bigr)\Bigr]
\end{multline}
with $\widehat{T}=1-T/T_\mathrm{c}$, where $P_\mathrm{c}=22.064\,\mathrm{MPa}$ and $T_\mathrm{c}=647.096\,\mathrm{K}$ are the critical pressure and temperature of water, respectively.

Futerko \& Hsing have initiated the semi-empirical modeling of vapor sorption in Nafion. Several of the developed models are based on modified Flory--Huggins solution theory \cite{futerko:99,choi:05a,kusoglu:11}, or multilayer adsorption theory, in its finite-layer Brunauer--Emmett--Teller (BET) form \cite{thampan:00,yang:04b,takata:07}, in the infinite-layer limit \cite{freger:97}, as a superposition of different sorption modes \cite{shi:16}, or extended by elastic swelling \cite{choi:03,kusoglu:09}. Other swelling models use thermal, chemical and elastic equilibrium assumptions \cite{eikerling:11,kreuer:13}. Many of these sorption models are based upon an additive decomposition of the total water content into a chemically bound (subscript b in Tab.~\ref{tab:lambda_v}) or clustered (c) part and a free (f) part.

Choi et al.~\cite{choi:05a} proposed a model in which they combine Flory--Huggins solution theory with elastic polymer swelling, using BET theory for the strongly bound water molecules. However, their description of the Flory--Huggins interaction parameter $\chi$ isn't explicit enough to allow us to reproduce their general sorption isotherm. Kreuer \cite{kreuer:13} proposed a thermodynamic sorption model with account for internal elastic pressure, which, after correction of the model equations \cite{kreuer:18}, underestimates the membrane hydration due to the treatment of Nafion as an elastomer. These models are therefore excluded from our following uncertainty analysis.

In Fig.~\ref{fig:lambda_v}b the sorption models are plotted at $30\,^\circ\mathrm{C}$ (where possible) to reveal their temperature-independent scatter, which is of an extent comparable to the polynomial correlations. Models that include a temperature dependence are also plotted in Fig.~\ref{fig:lambda_v}c at two different temperatures, highlighting that the effect of temperature is still far from understood. Four out of six models predict an increase in water uptake at high activity with rising temperature, whereas the other two predict the opposite. The experimental data these models have been validated on, as well as the proposed polynomial expressions, likewise disagree on the effect of temperature.

Thampan et al.~\cite{thampan:00} first recognized that finite-layer BET theory can be used to fit experimental vapor uptake data very well. Although Costamagna et al.~\cite{costamagna:08} and Li et al.~\cite{li:13} later demonstrated that the Guggenheim--Anderson--de Boer (GAB) equation and a dual-mode (DM) model work just as well, the BET isotherm presents a good compromise between physical interpretation, high quality of fit, and suitability for model implementation (being an explicit relationship between $a$ and $\lambda_\mathrm{v}$, unlike the more complex implicit sorption models):
\begin{equation}
\label{eq:bet}
\lambda_\mathrm{v} = \lambda_\mathrm{m}\frac{Ka}{1-a}\frac{1-(n+1)a^n+n a^{n+1}}{1+(K-1)a-Ka^{n+1}}
\end{equation}
where $K$ denotes the ratio of the absorption equilibrium constant of the first layer to that of the subsequent layers, determining the shape of $\lambda_\mathrm{v}$ at low relative humidity. $n$ is the number of adsorbed layers, governing the increase of water uptake at high relative humidity. The water loading at monolayer coverage $\lambda_\mathrm{m}$ can be estimated by \cite{thampan:00}
\begin{equation}
\lambda_\mathrm{m} = \frac{a_\mathrm{p}^\mathrm{PEM}}{A_\mathrm{w}}\frac{V_\mathrm{m}}{N_\mathrm{A}}
\end{equation}
where the area occupied by each adsorbed water molecule on the pore surface is approximately given by \cite{emmett:37}
\begin{equation}
A_\mathrm{w} = \sqrt{3}\left(\frac{V_\mathrm{w}}{2N_\mathrm{A}}\right)^{2/3}.
\end{equation}
$N_\mathrm{A}$ is the Avogadro constant and $a_\mathrm{p}^\mathrm{PEM}=210\,\mathrm{m}^2\,\mathrm{cm}^{-3}$ the pore surface area per unit volume of the membrane \cite{divisek:98}. Fitting Eq.~\ref{eq:bet} to published experiments on Nafion 117, NR211 and NR212 at temperatures between $20^\circ\mathrm{C}$ and $30^\circ\mathrm{C}$ \cite{springer:91,morris:93,rivin:01,jalani:05a,takata:07,peron:10,kusoglu:12} yields $K=92$ and $n=12.8$, which is close to the finding of Thampan et al.~\cite{thampan:00}. We use this BET isotherm for the baseline model parameterization. It also provides convenient oversaturation behavior, as depicted in Fig.~\ref{fig:lambda_v}d: As $a\to\infty$, $\lambda_\mathrm{v}\to\lambda_\mathrm{m}n\approx23$, which happens to coincide with the reported hydration number for liquid-equilibrated Nafion membranes. We therefore use $\lambda_\mathrm{l}=\lambda_\mathrm{m}n$ in Eqs.~\ref{eq:xi}, \ref{eq:xi_multilinear}, \ref{eq:DT} and \ref{eq:lambda_eq}.

\subsection{Gas diffusivity}
\label{sec:gas_diff}

An accurate way to estimate the effective binary diffusion coefficients $\mathcal{D}_{X,Y}$ for Eq.~\ref{eq:maxwell_stefan} is given by Chapman--Enskog kinetic gas theory \cite{bird:02,green:08}. Assuming the ideal gas law, they can be calculated as
\begin{equation}
\label{eq:chapman_enskog}
\mathcal{D}_{X,Y} = M_\mathrm{p}\frac{3}{8}\sqrt{\frac{RT}{2\pi}\left(\frac{1}{m_X}+\frac{1}{m_Y}\right)}\frac{k_\mathrm{B}T}{P\sigma^2_{X,Y}\Omega_{X,Y}}
\end{equation}
where $k_\mathrm{B}$ is the Boltzmann constant, $m_X$ denotes the molar mass of substance $X$, and
\begin{equation}
M_\mathrm{p}=\frac{\epsilon_\mathrm{p}}{\tau_\mathrm{p}^2}(1-s)^\phi
\end{equation}
is the microstructure factor of the pores, with compression-dependent porosity $\epsilon_\mathrm{p}$ and pore tortuosity $\tau_\mathrm{p}$ as detailed in Sec.~\ref{sec:compression}. For the saturation exponent $\phi$, we use $\phi^\mathrm{GDL}=3$ \cite{hwang:12} and $\phi^\mathrm{CL}=1.5$ \cite{fathi:17}. A curve fit of the collision integral $\Omega_{X,Y}$ is given by \cite{neufeld:72}
\begin{multline}
\Omega_{X,Y} = \frac{1.06036}{(T^*)^{0.15610}} + \frac{0.19300}{\exp\left[0.47635T^*\right]} \\+ \frac{1.03587}{\exp\left[1.52996T^*\right]} + \frac{1.76474}{\exp\left[3.89411T^*\right]}
\end{multline}
where $T^*=k_\mathrm{B}T/\varepsilon_{X,Y}$. For non-polar gases, the combining rules $\sigma_{X,Y}=(\sigma_X+\sigma_Y)/2$ and $\varepsilon_{X,Y}=\sqrt{\varepsilon_X\varepsilon_Y}$ can be used, where $\sigma_X$ and $\varepsilon_X$ are the Lennard--Jones collision diameters and potential depths, respectively, which are tabulated in the literature \cite{hirschfelder:67,berendsen:81,bird:02}. For inter-diffusion of a polar and a non-polar gas, modified theories with increased complexity exist, such as the one by Brokaw \cite{brokaw:69}. For simplicity it is assumed here that Chapman--Enskog theory also applies with sufficient accuracy for humid gases, i.e., that the dipole moment of water molecules can be neglected.

Assuming cylindrical pores with effective average pore radius $r_\mathrm{p}$, the Knudsen diffusivities $D_{\mathrm{K},X}$ are given by \cite{knudsen:09}
\begin{equation}
D_{\mathrm{K},X} = \frac{8r_\mathrm{p}}{3}\sqrt{\frac{RT}{2\pi m_X}}.
\end{equation}
To account for pore narrowing by liquid water, we adopt the quadratic law $r_\mathrm{p} = r_\mathrm{p,dry} s_\mathrm{w}^2$, which fits the data by Hutzenlaub et al.~\cite{hutzenlaub:13} well, where $s_\mathrm{w}$ is the reduced wetting phase saturation as defined in Eq.~\ref{eq:dpcds}. The following dry radii are used: $r_\mathrm{p,dry}^\mathrm{GDL}=15\,\upmu\mathrm{m}$ for a SGL 24 BC \cite{wood:06} and $r_\mathrm{p,dry}^\mathrm{CL}=20\,\mathrm{nm}$ for a Nafion/carbon black CL \cite{ono:13}.

\subsection{Liquid water transport}
\label{sec:liquid_water}

The effective liquid water transport coefficient $D_s$ in Eq.~\ref{eq:j_s} depends on several material properties. In particular the functional relationship between saturation $s$ and capillary pressure $p_\mathrm{c}$ is a topic of extensive research with a large variety of fitting functions that have been proposed \cite{si:15}. It strongly depends not only on the wettability and compression of the porous medium, but also on the exact kind of water transport process (primary injection, withdrawal, further injections). We found the overall impact of $D_s$ on fuel cell performance to be mostly small, though, as will be shown in Part II of this series. Therefore, we focus on a single parameter set here rather than screening the literature for different parameterizations. A study on the effects of artificially altered capillary pressure--saturation relationships can be found in \cite{wang:08b}. The van Genuchten law, which Gostick et al.~\cite{gostick:09} have found to apply to many common GDLs, is used here:
\begin{equation}
\label{eq:dpcds}
\frac{\partial p_\mathrm{c}}{\partial s} = \frac{p_\mathrm{b}}{lm}\left(s_\mathrm{w}^{-1/m}-1\right)^{1/l-1}s_\mathrm{w}^{-1/m-1},\quad s_\mathrm{w}=\frac{1-s}{1-s_\mathrm{im}}
\end{equation}
with the following parameters for the secondary water injection curve of compressed SGL carbon paper: $m=0.6$, $l=100$, a breakthrough pressure of $p_\mathrm{b}=1.07\,\mathrm{bar}$, and an immobile saturation of $s_\mathrm{im}=0.08$ \cite{gostick:09}. Mualem's model is used for the relative hydraulic permeability, reading \cite{zamel:11}
\begin{equation}
K_\mathrm{rel} = \left(1-s_\mathrm{w}\right)^2\left(1-s_\mathrm{w}^{1/m}\right)^{2m}+10^{-6}
\end{equation}
where the small offset serves to bypass numerical difficulties under dry conditions, i.e., to avoid that $K_\mathrm{rel}\to0$ as $s\to s_\mathrm{im}$. We note, however, that the relative permeability in partially saturated GDLs is an active area of research, and alternate expressions such as S-shaped functions were recently proposed \cite{holzer:17b}. For the absolute permeability of the GDLs, the semi-heuristic Carman--Kozeny equation for fibrous porous media is employed, as it proved to work well for carbon paper \cite{gostick:06}:
\begin{equation}
K_\mathrm{abs}^\mathrm{GDL} = \frac{\epsilon_\mathrm{p}^3 d_\mathrm{f}^2}{16k_\mathrm{K}(1-\epsilon_\mathrm{p})^2}
\end{equation}
where $d_\mathrm{f}=8.0\,\upmu\mathrm{m}$ and $k_\mathrm{K}=4.54$ are the fiber diameter and the Kozeny constant for a SGL 24 BA \cite{gostick:06}. A constant value of $K_\mathrm{abs}^\mathrm{CL}=0.1\,\upmu\mathrm{m}^2$ \cite{yi:99} is assumed for the CLs. Finally, to complete the definition of the effective liquid water transport coefficient $D_s$ in Eq.~\ref{eq:j_s}, the internationally recommended correlation of the dynamic viscosity of liquid water at $1\,\mathrm{bar}$ up to $110\,^\circ\mathrm{C}$ is used \cite{huber:09}:
\begin{multline}
\mu=\Big(280.68\overline{T}^{-1.9} + 511.45\overline{T}^{-7.7} \\+ 61.131\overline{T}^{-19.6} + 0.45903\overline{T}^{-40}\Big)\,\upmu\mathrm{Pa\,s}
\end{multline}
with $\overline{T}=T/300\,\mathrm{K}$. In the molar volume of liquid water $V_\mathrm{w}=m_\mathrm{w}/\rho_\mathrm{w}$, $m_\mathrm{w}=18.015\,\mathrm{g\,mol}^{-1}$ is the molar mass of water, and $\rho_\mathrm{w}$ its mass density given at standard atmospheric pressure by \cite{wagner:93}
\begin{multline}
\rho_\mathrm{w} = \rho_\mathrm{c}\Bigl(1+1.9927\widehat{T}^{1/3}+1.0997\widehat{T}^{2/3}-0.51084\widehat{T}^{5/3}\\-1.7549\widehat{T}^{16/3}-45.517\widehat{T}^{43/3}-674694\widehat{T}^{110/3}\Bigr)
\end{multline}
with $\widehat{T}$ and $\rho_\mathrm{c}$ like in Eq.~\ref{eq:Psat}.

\subsection{Vapor sorption}
\label{sec:vapor_sorption}

\begin{table*}
	\centering
	\caption{Review of mass transfer coefficients for water vapor absorption/desorption in Nafion membranes.}
	\label{tab:ad_coeff}
	\scalebox{\tabscale}{\begin{tabular}{lllll}
	\toprule
	Publication & Mass transfer coefficients [$10^{-3}\,\mathrm{cm\,s}^{-1}$] & Temperature & Activation energy & Membrane\\
	\midrule[\heavyrulewidth]
	Rivin et al., 2001 \cite{rivin:01} & $1031$ & 20 or 32\,$^\circ$C & & N117\\\midrule
	Ye \& LeVan, 2003 \cite{ye:03} & $\sim p_{\mathrm{H}_2\mathrm{O}}^{1.5}$ & 23.5--25\,$^\circ$C\\\midrule
	Berg et al., 2004 \cite{berg:04} & $0.57$ & 70--80\,$^\circ$C & & N112\\\midrule
	\multirow{2}{*}{Ge et al., 2005 \cite{ge:05}} & $k_\mathrm{a}=1.14f_\mathrm{w}$ & \multirow{2}{*}{30\,$^\circ$C} & \multirow{2}{*}{$20\,\mathrm{kJ\,mol}^{-1}$} & \multirow{2}{*}{N112,115,117}\\
	& $k_\mathrm{d}=4.59f_\mathrm{w}$\\\midrule
	Satterfield et al., 2008 \cite{satterfield:08} & $k_\mathrm{d}=\text{0.14--0.29}$ & 70\,$^\circ$C & 25--31\,$\mathrm{kJ\,mol}^{-1}$ & N112,115,1110,1123\\\midrule
	Monroe et al., 2008 \cite{monroe:08} & $630$ & 50\,$^\circ$C & & N112,115,117\\\midrule
	Hallinan \& Elabd, 2009 \cite{hallinan:09} & $20$ & 30\,$^\circ$C & & N117\\\midrule
	Adachi et al., 2010 \cite{adachi:10,kongkanand:11} & $0.45$ & 70\,$^\circ$C & & N112,115,117, DE2021CS\\\midrule
	He et al., 2011$^*$ \cite{he:11} & $1.85\max\{0,\lambda-\lambda_0\}^{1.25}$, $\lambda_0=3.17$ & 25\,$^\circ$C & & NR212\\\midrule
	Tabuchi et al., 2011 \cite{tabuchi:11} & $k_\mathrm{d}=\text{0.2--1}$ & 30\,$^\circ$C & & N1110\\\midrule
	\multirow{2}{*}{Kongkanand, 2011 \cite{kongkanand:11}} & $k_\mathrm{a}=0.0184a^2+0.0586a+0.129$ & \multirow{2}{*}{80\,$^\circ$C} & $28.1\,\mathrm{kJ\,mol}^{-1}$ & \multirow{2}{*}{DE2020}\\
	& $k_\mathrm{d}=0.256\phantom{0}a^2+0.148\phantom{0}a+0.191$ & & $29.7\,\mathrm{kJ\,mol}^{-1}$\\\midrule
	Kusoglu \& Weber, 2012$^*$ \cite{kusoglu:12b,kientiz:11} & $0.68\max\{0,\lambda-\lambda_0\}^{1.6}$ & 25--80\,$^\circ$C & & NR211,212\\\midrule
	Didierjean et al., 2015$^*$ \cite{didierjean:15} & $400V_\mathrm{m}P_\mathrm{sat}/RT$ & 25\,$^\circ$C & & N117\\
	\midrule[\heavyrulewidth]
	\multicolumn{5}{@{}l@{}}{$^*$To be divided by the sorption slope $\partial\lambda_\mathrm{v}/\partial a$ when used in Eq.~\ref{eq:Sad}.}\\
	\end{tabular}}
\end{table*}

\begin{figure}
	\centering
	\includegraphics{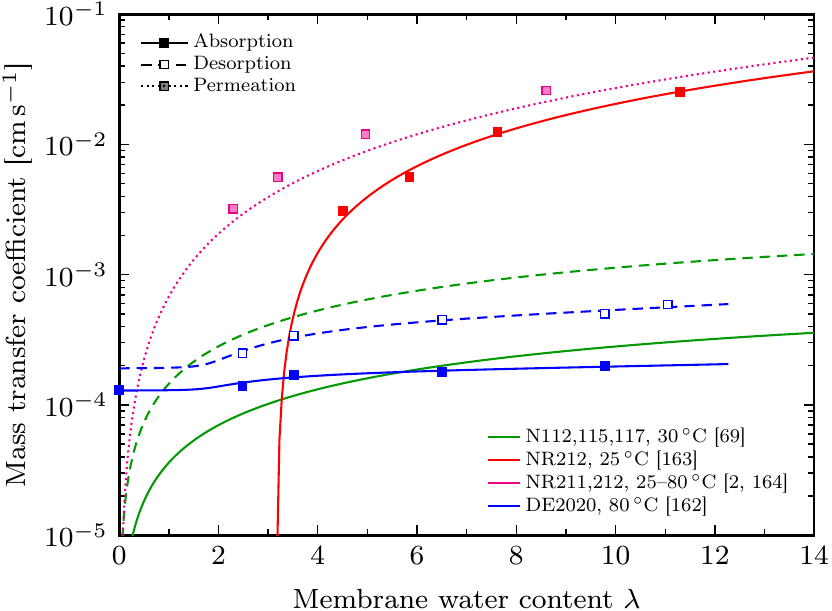}
	\caption{Experimentally determined interfacial mass transfer coefficients for Nafion membranes as a function of water content. Individual symbols represent measurement data and lines show fitted functions. Data from \cite{kongkanand:11,kusoglu:12b} were converted from relative humidity to water content using Eq.~\ref{eq:bet}.}
	\label{fig:ad_coeff}
\end{figure}

Numerous experiments have been carried out to measure the interfacial mass transfer coefficients $k_\mathrm{a}$ and $k_\mathrm{d}$ in Eq.~\ref{eq:Sad}, as listed in Tab.~\ref{tab:ad_coeff}. They differ in the experimental procedure (absorption, desorption, permeation, liquid vs.\ vapor basins) and the driving force considered (concentration or hydration number gradient vs.\ activity gradient). Measurements that are based upon a water activity difference between the ionomer ($a_\mathrm{i}$) and vapor phases ($a$) are marked with an asterisk in Tab.~\ref{tab:ad_coeff}. These coefficients can be approximately converted to $k_\mathrm{a}$ and $k_\mathrm{d}$ for use in Eq.~\ref{eq:Sad} by dividing them by the local slope of the sorption isotherm, $\partial\lambda_\mathrm{v}/\partial a$ \cite{didierjean:15}, because for small deviations from equilibrium,
\begin{equation}
\label{eq:k_ad_conversion}
(\lambda-\lambda_\mathrm{v}) \approx \frac{\partial\lambda_\mathrm{v}}{\partial a}(a_\mathrm{i}-a).
\end{equation}

Some of the more systematic studies that include the moisture dependence of the water transport resistance at the ionomer--gas interface are compared in Fig.~\ref{fig:ad_coeff}. Ye \& LeVan \cite{ye:03} reported an approximate power-law correlation with the partial pressure of water vapor. Since it is unclear how their expression translates to Eq.~\ref{eq:Sad} though, we exclude this early result from the following analysis. Ge et al.~\cite{ge:05} proposed to take $k_\mathrm{a}$ and $k_\mathrm{d}$ proportional to the ionomer's water volume fraction $f_\mathrm{w}$ analogous to the intra-diffusion coefficient of water in bulk Nafion as proposed by Weber \& Newman \cite{weber:04a} (cf.\ Tab.~\ref{tab:D_lambda}). They also adopted the activation energy $E_\mathrm{ad}=20\,\mathrm{kJ\,mol}^{-1}$ from a diffusion measurement by Yeo \& Eisenberg \cite{yeo:77}. Later measurements of the activation energy of sorption have yielded larger values, $E_\mathrm{ad}\approx29\,\mathrm{kJ\,mol}^{-1}$ \cite{majsztrik:07,satterfield:08,kongkanand:11}. He et al.~\cite{he:11} and Kusoglu et al.~\cite{kusoglu:12b} proposed power laws in $\lambda$ to express the mass transfer coefficients, based on data that possibly stem from the same measurement carried out at the Lawrence Berkeley National Laboratory, but assuming different residual hydration numbers $\lambda_0$ (3.17 vs.\ 0 in Fig.~\ref{fig:ad_coeff}). Converting their measured data points from activity-driven to hydration number-driven using Eq.~\ref{eq:k_ad_conversion} yields an approximately constant mass transfer coefficient $k_\mathrm{a,d}\approx10^{-3}\,\mathrm{cm\,s}^{-1}$ independent of $\lambda$ (data not shown). Kongkanand \cite{kongkanand:11} used a polynomial in water activity to fit the coefficients.

Although there is disagreement on whether the interfacial mass transfer grows with increasing $\lambda$ \cite{ye:03,ge:05,kongkanand:11} or not \cite{he:11,kusoglu:12b}, the studies that differentiate between absorption and desorption agree that the former is substantially slower than the latter \cite{ge:05,kongkanand:11,didierjean:15}. To date, the true dependence on moisture is essentially an open problem. Most researchers report the mass transfer coefficients in the range $10^{-4}$--$10^{-2}\,\mathrm{cm\,s}^{-1}$ in the relevant temperature range of $50$--$90\,^\circ\mathrm{C}$. This is also the range numerically examined in early models by Okada \cite{okada:98,okada:99}. With these considerations in mind, we adopt Ge's correlation for the baseline simulation.

\subsection{Evaporation and condensation}
\label{sec:evap_cond}

\begin{table*}
	\centering
	\caption{Review of evaporation/condensation rates.}
	\label{tab:ec_rates}
	\scalebox{\tabscale}{\begin{tabular}{llll}
	\toprule
	Publication & Rate expressions & Coefficients & Rate values$^*$\\\midrule[\heavyrulewidth]
	Nguyen \& White, 1993$^\dagger$ \cite{nguyen:93} & $\gamma_\mathrm{c}=k_\mathrm{c}$ & $k_\mathrm{c}=1\,\mathrm{s}^{-1}$ & $\gamma_\mathrm{c}=1\,\mathrm{s}^{-1}$\\
	& $\gamma_\mathrm{e}=k_\mathrm{c}$ & & $\gamma_\mathrm{e}=1\,\mathrm{s}^{-1}$\\\midrule
	Nguyen, 1999 \cite{nguyen:99} & $\gamma_\mathrm{c}=k_\mathrm{c}\epsilon_\mathrm{p}(1-s)$ & $k_\mathrm{c}=100\,\mathrm{s}^{-1}$ \cite{nguyen:99} & $\gamma_\mathrm{c}=56\,\mathrm{s}^{-1}$\\
	& $\gamma_\mathrm{e}=k_\mathrm{e}\epsilon_\mathrm{p}sRT/V_\mathrm{w}$ & $k_\mathrm{e}=100\,\mathrm{atm}^{-1}\,\mathrm{s}^{-1}$ \cite{nguyen:99} & $\gamma_\mathrm{e}=22\,000\,\mathrm{s}^{-1}$\\
	& & $k_\mathrm{c}=100\,\mathrm{s}^{-1}$ \cite{song:06} & $\gamma_\mathrm{c}=56\,\mathrm{s}^{-1}$\\
	& & $k_\mathrm{e}=1\,\mathrm{atm}^{-1}\,\mathrm{s}^{-1}$ \cite{song:06} & $\gamma_\mathrm{e}=220\,\mathrm{s}^{-1}$\\\midrule
	He et al., 2000 \cite{he:00} & $\gamma_\mathrm{c}=k_\mathrm{c}\epsilon_\mathrm{p}(1-s)y_{\mathrm{H}_2\mathrm{O}}$ & $k_\mathrm{c}=100\,\mathrm{s}^{-1}$ \cite{he:00} & $\gamma_\mathrm{c}=14\,\mathrm{s}^{-1}$\\
	& $\gamma_\mathrm{e}=k_\mathrm{e}\epsilon_\mathrm{p}sRT/V_\mathrm{w}$ & $k_\mathrm{e}=100\,\mathrm{atm}^{-1}\,\mathrm{s}^{-1}$ \cite{he:00} & $\gamma_\mathrm{e}=22\,000\,\mathrm{s}^{-1}$\\
	& & $k_\mathrm{c}=5000\,\mathrm{s}^{-1}$ \cite{meng:07} & $\gamma_\mathrm{c}=700\,\mathrm{s}^{-1}$\\
	& & $k_\mathrm{e}=10^{-4}\,\mathrm{Pa}^{-1}\,\mathrm{s}^{-1}$ \cite{meng:07} & $\gamma_\mathrm{e}=1200\,\mathrm{s}^{-1}$\\
	& & $k_\mathrm{c}=100\,\mathrm{s}^{-1}$ \cite{nguyen:10} & $\gamma_\mathrm{c}=14\,\mathrm{s}^{-1}$\\
	& & $k_\mathrm{e}=5\,\mathrm{atm}^{-1}\,\mathrm{s}^{-1}$ \cite{nguyen:10} & $\gamma_\mathrm{e}=1100\,\mathrm{s}^{-1}$\\\midrule
	Natarajan \& Nguyen, 2001 \cite{natarajan:01} & $\gamma_\mathrm{c}=k_\mathrm{c}\epsilon_\mathrm{p}(1-s)RTy_{\mathrm{H}_2\mathrm{O}}$ & $k_\mathrm{c}=\text{N/A}$ & $\gamma_\mathrm{c}=\text{N/A}$\\
	& $\gamma_\mathrm{e}=k_\mathrm{e}\epsilon_\mathrm{p}sRT/V_\mathrm{w}$ & $k_\mathrm{e}=\text{N/A}$ & $\gamma_\mathrm{e}=\text{N/A}$\\\midrule
	Nam \& Kaviany, 2003 \cite{nam:03} & $\gamma_\mathrm{c}=\Gamma_\mathrm{m}a_\mathrm{lg}\sqrt{RT/2\pi m_\mathrm{w}}$ & $\Gamma_\mathrm{m}=0.006$ & $\gamma_\mathrm{c}=970\,\mathrm{s}^{-1}$\\
	& $\gamma_\mathrm{e}=\Gamma_\mathrm{m}a_\mathrm{lg}\sqrt{RT/2\pi m_\mathrm{w}}$ & $a_\mathrm{lg}=1000\,\mathrm{m}^{-1}$ & $\gamma_\mathrm{e}=970\,\mathrm{s}^{-1}$\\\midrule
	Weber et al., 2004 \cite{weber:04d} & $\gamma_\mathrm{c}=k_\mathrm{m}a_\mathrm{lg}RT$ & $k_\mathrm{m}a_\mathrm{lg}=100\,\mathrm{mol}\,\mathrm{bar}^{-1}\,\mathrm{cm}^{-3}\,\mathrm{s}^{-1}$ & $\gamma_\mathrm{c}=2\,900\,000\,\mathrm{s}^{-1}$\\
	& $\gamma_\mathrm{e}=k_\mathrm{m}a_\mathrm{lg}RT$ & & $\gamma_\mathrm{e}=2\,900\,000\,\mathrm{s}^{-1}$\\\midrule
	Birgersson et al.,  2005 \cite{birgersson:05} & $\gamma_\mathrm{c}=k_\mathrm{c}\epsilon_\mathrm{p}$ & $k_\mathrm{c}=100\,\mathrm{s}^{-1}$ & $\gamma_\mathrm{c}=70\,\mathrm{s}^{-1}$\\
	& $\gamma_\mathrm{e}=k_\mathrm{e}sRT/m_\mathrm{w}$ & $k_\mathrm{e}=100\,\mathrm{s}\,\mathrm{m}^{-2}$ & $\gamma_\mathrm{e}=3\,300\,000\,\mathrm{s}^{-1}$\\\midrule
	Eikerling et al., 2006$^\ddagger$ \cite{eikerling:06} & $\gamma_\mathrm{e}=k_\mathrm{e}RT\xi_\mathrm{lg}/N_\mathrm{A}L^\mathrm{CL}$ & $k_\mathrm{e}=1.4\!\times\!10^{18}\,\mathrm{atm}^{-1}\,\mathrm{cm}^{-2}\,\mathrm{s}^{-1}$ & $\gamma_\mathrm{e}=13\,000\,\mathrm{s}^{-1}$\\\midrule
	Wu et al., 2009 \cite{wu:09b} & $\gamma_\mathrm{c}=\Gamma_\mathrm{m}\Gamma_\mathrm{s}a_\mathrm{p}(1-s)\sqrt{RT/2\pi m_\mathrm{w}}$ & $\Gamma_\mathrm{m}=0.006$ & $\gamma_\mathrm{c}=1\,500\,000\,\mathrm{s}^{-1}$\\
	& $\gamma_\mathrm{e}=\Gamma_\mathrm{m}\Gamma_\mathrm{s}a_\mathrm{p}s\sqrt{RT/2\pi m_\mathrm{w}}$ & $\Gamma_\mathrm{s}\lesssim0.2$ & $\gamma_\mathrm{e}=390\,000\,\mathrm{s}^{-1}$\\
	\midrule[\heavyrulewidth]
	\multicolumn{4}{@{}l@{}}{$^*$Evaluated at $T=80\,^\circ\mathrm{C}$, $\epsilon_\mathrm{p}=0.7$, $s=0.2$, $y_{\mathrm{H}_2\mathrm{O}}=0.25$, $\Gamma_\mathrm{s}=0.1$, $a_\mathrm{p}=20\,\mathrm{m}^2\,\mathrm{cm}^{-3}$, $L^\mathrm{CL}=10\,\text{\textmu}\mathrm{m}$, $\xi_\mathrm{lg}=200$.}\\
	\multicolumn{4}{@{}l@{}}{$^\dagger$Evaporation/condensation in gas channel.}\\
	\multicolumn{4}{@{}l@{}}{$^\ddagger$Evaporation model for cathode catalyst layer only.}\\
	\end{tabular}}
\end{table*}

\begin{figure}
	\centering
	\includegraphics{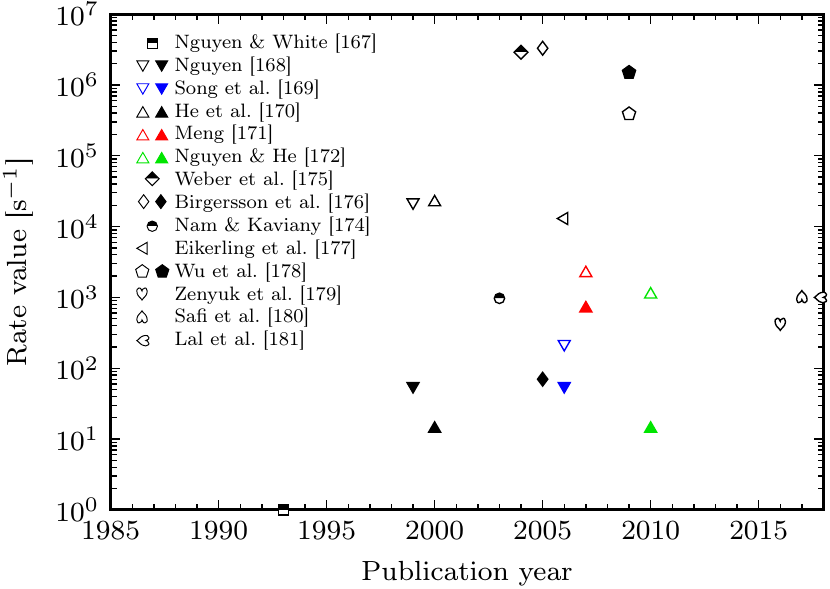}
	\caption{Review of evaporation rates (open symbols) and condensation rates (closed symbols). Symbols are half-filled if both are equal. Colored symbols denote modified parameters used in conjunction with earlier established rate expressions (black symbols) as listed in Tab.~\ref{tab:ec_rates}. For comparison, a recent series of \textit{ex-situ} and \textit{in-situ} evaporation measurements in SGL 24 BA \cite{zenyuk:16,safi:17} and Toray TGP-H-060 \cite{lal:18} were recast into the form of Eq.~\ref{eq:Sec} and shown as heart symbols.}
	\label{fig:ec_rates}
\end{figure}

Eq.~\ref{eq:Sec} is the commonly employed out-of-equilibrium way to account for liquid--vapor phase change in macro-ho\-mo\-ge\-ne\-ous PEMFC modeling with explicit representation of both phases. A comprehensive comparison of different expressions for the rates $\gamma_\mathrm{e,c}$ developed in modeling is given in Tab.~\ref{tab:ec_rates}. In the final column, they are evaluated for a typical state of fuel cell operation, demonstrating that the effective evaporation and condensation rates used in published two-phase models vary over as much as five orders of magnitude. Fig.~\ref{fig:ec_rates} shows the timeline of when these rates came about in the literature, revealing no sign of convergence nor even a trend.

A feature shared by many models is that the condensation (evaporation) rate is assumed to be proportional to $1-s$ ($s$) to account for the change of available phase boundary, as first proposed by Nguyen \cite{nguyen:99}. He et al.~\cite{he:00} proposed that condensation be proportional to the molar water vapor fraction $y_{\mathrm{H}_2\mathrm{O}}$, but more recently developed models have abandoned this assumption. Nam \& Kaviany \cite{nam:03} suggested to use the Hertz--Knudsen equation from kinetic gas theory to model both directions of the phase change, assuming continuity of the temperature at the phase boundary and equal rates. Weber et al.~\cite{weber:04d} used a simpler expression, also based on equal rates for evaporation and condensation. These formulas explicitly include the liquid--gas interfacial area density $a_\mathrm{lg}$, which is unknown a priori, making even an estimation of the rates difficult. Wu et al.~\cite{wu:09b} estimate it to be $a_\mathrm{lg}=\Gamma_\mathrm{s}a_\mathrm{p}(1-s)$ for condensation and $a_\mathrm{lg}=\Gamma_\mathrm{s}a_\mathrm{p}s$ for evaporation with an interfacial area accommodation coefficient $\Gamma_\mathrm{s}\lesssim0.2$. $a_\mathrm{p}$ denotes the average pore surface area density of the porous domain. A recent experimental study \cite{zenyuk:16} has shown that $a_\mathrm{lg}\sim s$ and that there is no clear correlation between the evaporation rate per interfacial area and $s$, suggesting that the evaporation rate should indeed increase linearly with $s$.

It is known from numerous experimental measurements that the condensation coefficient of water is roughly a decade larger than the evaporation coefficient, and that both decrease with increasing pressure and temperature \cite{marek:01}. The proposed coefficient $\Gamma_\mathrm{m}=0.006$ in Nam's and Wu's models is consistent with this experimental data at atmospheric pressure for condensation, but about a decade too large for evaporation. For this reason we implement the Hertz--Knudsen law in the form proposed by Wu et al.\ (final row in Tab.~\ref{tab:ec_rates}) as the baseline parameterization of phase change with $\Gamma_\mathrm{s}=0.1$, $\Gamma_\mathrm{m}=0.006$ for condensation, but $\Gamma_\mathrm{m}=0.0005$ for evaporation, analogous to \cite{vetter:18}. We also use $s_\mathrm{nw}$ in place of $s$ to prevent $s_\mathrm{nw}$ from becoming negative during evaporation. The specific pore surface area $a_\mathrm{p}$ is modeled as a function of compression as discussed in Sec.~\ref{sec:compression}.

It should be noted, however, that the general validity of the Hertz--Knudsen equation was recently questioned and that it was hypothesized that evaporation/condensation be governed by a balance of tiny pressure differences and momentum flux instead \cite{holyst:15}. The Hertz--Knudsen rates used here are among the higher ones used in modeling works (cf.\ Fig.~\ref{fig:ec_rates}), quickly yielding saturated vapor where liquid water is present. In this quasi-equilibrium regime, the exact values of the rates become irrelevant \cite{weber:04c,basu:09}, and any functional dependencies of the rate expressions on $s$, $T$ etc.\ can essentially be dropped. This is in agreement with the reported observation that evaporation in partially saturated GDLs is fast enough to be transport-limited (diffusion-limited) rather than area- or kinetics-limited \cite{zenyuk:16,safi:17,lal:18}.

\subsection{Latent heat}
\label{sec:latent_heat}

The molar latent heat of water condensation $H_\mathrm{ec}$ used in Eq.~\ref{eq:ST} can be parameterized as
\begin{equation}
\label{eq:Hec}
H_\mathrm{ec} = 52.51\exp\left[0.261\widetilde{T}-0.044\widetilde{T}^2-0.0044\widetilde{T}^3\right]\,\mathrm{kJ\,mol}^{-1}
\end{equation}
with $\widetilde{T}=\ln\left[1-T/T_\mathrm{c}\right]$. This is a global least-squares fit to the tabulated data from \cite{haynes:16}. The latent heat of water sorption can be written as
\begin{equation}
\label{eq:Had}
H_\mathrm{ad} = H_\mathrm{ec} + H_\mathrm{mix}
\end{equation}
where the mixing enthalpy $H_\mathrm{mix}$, which is often assumed constant or even neglected in published models, is a function of membrane hydration \cite{ostrovskii:96,reucroft:02} and temperature \cite{wadso:13} for Nafion. Since a reliable parameterization for $H_\mathrm{mix}$ is absent from the literature, a new one is proposed here, based on the measurements on Nafion 115 by Wads\"{o} \& Jannasch \cite{wadso:13}. Their data indicate that $H_\mathrm{mix}$ essentially decays exponentially in $\lambda$, with a temperature-dependent deviation in the very dry regime ($\lambda<3$). We fitted the phenomenological approximation
\begin{equation}
\label{eq:Hmix}
H_\mathrm{mix} = \left(a_1\exp\left[-b_1\lambda\right] + a_2\lambda\exp\left[-b_2\lambda^2\right]\right)\,\mathrm{kJ\,mol}^{-1}
\end{equation}
with temperature-dependent coefficients to their data in the range $40\,^\circ\mathrm{C}\leq T\leq100\,^\circ\mathrm{C}$ and $0.05\leq\lambda\leq5$ and obtained least squared residuals for
\begin{equation}
\begin{split}
a_1 &= -107.5\overline{T}^2+253.9\overline{T}-138.7\\
a_2 &= 106.8\overline{T}-102.4\\
b_1 &= 2.006\overline{T}^2-4.365\overline{T}+2.931\\
b_2 &= 108.7\overline{T}^2-262.8\overline{T}+159.5\\
\end{split}
\end{equation}
where $\overline{T}=T/300\,\mathrm{K}$.

\subsection{Compression and contact resistivities}
\label{sec:compression}

To account for the effects of clamping pressure on the computational domain, the layer thicknesses are modeled as a function of pressure. Given the compressive strain $\varepsilon_\mathrm{c}$, one can write
\begin{equation}
\label{eq:thickness}
L = L_0(1-\varepsilon_\mathrm{c})
\end{equation}
where $L_0$ denotes the thickness of the uncompressed layer. We model SGL 24 GDLs on both sides of the MEA, for which $L_0^\mathrm{GDL}=190\,\upmu\mathrm{m}$ \cite{sgl:09}. The relationship between strain and applied clamping pressure $P_\mathrm{cl}$ for these GDLs is \cite{kumbur:07b}
\begin{equation}
\varepsilon_\mathrm{c}^\mathrm{GDL} = -0.0083\left(\frac{P_\mathrm{cl}}{1\,\mathrm{MPa}}\right)^2+0.0911\left(\frac{P_\mathrm{cl}}{1\,\mathrm{MPa}}\right).
\end{equation}
The two catalyst layers are also compressed under applied pressure, but reliable measurement data is rare. We fitted the following two-parameter function to the compressive strain of the CL reported by Burheim et al.~\cite{burheim:14}:
\begin{equation}
\label{eq:eps_c_CL}
\varepsilon_\mathrm{c}^\mathrm{CL} = 0.422\left(1-\exp\left[-\frac{P_\mathrm{cl}}{0.970\,\mathrm{MPa}}\right]\right)
\end{equation}
Together with Eq.~\ref{eq:thickness}, this equation is used to express $L^\mathrm{CL}$ as a function of applied pressure. For the baseline simulation, $L_0^\mathrm{CL}=10\,\upmu\mathrm{m}$ and $P_\mathrm{cl}=1\,\mathrm{MPa}$ are used. Nafion NR211 with constant thickness $L_0^\mathrm{PEM}=25.4\,\upmu\mathrm{m}$ \cite{nafion:16} is chosen as the membrane, assuming for simplicity that swelling and compression cancel one another ($\varepsilon_\mathrm{c}^\mathrm{PEM}=0$).

Kumbur et al.~\cite{kumbur:07b} measured the average specific pore surface $a_\mathrm{p}$ at three different compaction pressures from $0$ to $1.4\,\mathrm{MPa}$ for a few GDLs from SGL Group with MPLs, finding a moderate increase with increasing pressure. Asymptotically, though, it is clear that at very large applied pressures, the pore surface area density must come down again. We therefore fitted a quadratic polynomial to their data for SGL 24 BC:
\begin{equation}
\label{eq:ap}
\frac{a_\mathrm{p}^\mathrm{GDL}}{1\,\mathrm{m}^2\,\mathrm{cm}^{-3}} = -1.96\left(\frac{P_\mathrm{cl}}{1\,\mathrm{MPa}}\right)^2 + 8.18\left(\frac{P_\mathrm{cl}}{1\,\mathrm{MPa}}\right) + 23.4.
\end{equation}
As no data seems to be available in the literature for CLs, we set $a_\mathrm{p}^\mathrm{CL}=a_\mathrm{p}^\mathrm{GDL}$.

Kumbur et al.\ also determined the compression dependence of porosity for SGL carbon paper (with MPL) in the pressure range up to 4\,MPa and proposed the relationship
\begin{equation}
\label{eq:eps_p_GDL}
\epsilon_\mathrm{p}^\mathrm{GDL} = \epsilon_{\mathrm{p},0}^\mathrm{GDL}\left(\frac{0.9}{1+\varepsilon_\mathrm{c}^\mathrm{GDL}}+0.1\right)
\end{equation}
where $\epsilon_{\mathrm{p},0}^\mathrm{GDL}=0.75$ \cite{kumbur:07b} is the porosity of the uncompressed GDL. While Eq.~\ref{eq:eps_c_CL} determines how much CLs are compressed, it is unclear how this changes their porosity. We assume that only the pore space of the CLs is compressed, which yields
\begin{equation}
\epsilon_\mathrm{p}^\mathrm{CL} = \frac{\epsilon_{\mathrm{p},0}^\mathrm{CL}-\varepsilon_\mathrm{c}^\mathrm{CL}}{1-\varepsilon_\mathrm{c}^\mathrm{CL}}
\end{equation}
where $\epsilon_{\mathrm{p},0}^\mathrm{CL}$ is the porosity of the uncompressed CL, taken as $0.4$ \cite{bernardi:92}. The relationship between trough-plane pore tortuosity and applied clamping pressure is commonly expressed indirectly, with tortuosity as a function of porosity. For the present model, a parabolic least-squares fit to experimental data for SGL 24 DA by Fl\"{u}ckiger et al.~\cite{flueckiger:08} is used,
\begin{equation}
\label{eq:tau_p}
\tau_\mathrm{p}^\mathrm{GDL} = -17.3\left(\epsilon_\mathrm{p}^\mathrm{GDL}\right)^2+18.8\epsilon_\mathrm{p}^\mathrm{GDL}-1.72,
\end{equation}
whereas the tortuosity of the CLs is assumed constant ($\tau_\mathrm{p}^\mathrm{CL}=1.5$ \cite{litster:13,babu:16}) due to the apparent absence of published measurement data on its compression dependence.

\begin{table}
	\centering
	\caption{Contact resistance parameters.}
	\label{tab:Rct}
	\scalebox{\tabscale}{\begin{tabular}{lcccccc}
	\toprule
	& \multicolumn{3}{c}{ECR} & \multicolumn{3}{c}{TCR}\\
	\cmidrule(lr){2-4}
	\cmidrule(lr){5-7}
	Interface & $R_0$ [$\mathrm{m}\Omega\,\mathrm{cm}^2$] & $\zeta$ [--] & Ref. & $R_0$ [$\mathrm{K}\,\mathrm{cm}^2\,\mathrm{W}^{-1}$] & $\zeta$ [--] & Ref.\\
	\midrule
	CL/GDL & $29.4$ & $0.89$ & \cite{nitta:08a} & $1.56$ & $0.71$ & \cite{khandelwal:06}\\
	GDL/BP & $3.34$ & $0.53$ & \cite{zhou:06} & $2.89$ & $0.64$ & \cite{sadeghifar:14}\\
	\bottomrule
	\end{tabular}}
\end{table}

To complete the constitutive parameterization of our model, the electrical and thermal contact resistivities $R_\mathrm{e}$ and $R_T$ in Eq.~\ref{eq:cr} remain to be specified for the different MEA interfaces. As recently highlighted \cite{vetter:17}, they usually follow power laws of the form
\begin{equation}
\label{eq:Rct}
R = R_0\left(\frac{P_\mathrm{cl}}{1\,\mathrm{MPa}}\right)^{-\zeta}
\end{equation}
where the coefficients $R_0$ and $\zeta$ vary with the materials considered -- a result that was also found theoretically for contacting fractal surfaces, for which the exponent is $\zeta\in[0.5,1]$ depending on the degree of material plasticity and surface roughness \cite{majumdar:91}. Numerical simulations based on contact mechanics have confirmed this relationship \cite{zhou:07,wu:08} for ECR at the GDL/BP interface. Thermal and electrical contact resistances are implemented in the present MEA model using the values listed in Tab.~\ref{tab:Rct}.

Finally, we note that through-plane TCR is significantly reduced by the presence of liquid water (e.g.,\ \cite{burheim:11b}). Since the data in the literature is too scattered though to allow for a reliable parameterization w.r.t.\ $s$, the dependency of the contact resistivities on moisture is neglected here. This aspect requires additional experimental clarification.

\section{Impact on fuel cell performance}
\label{sec:results}

\begin{table}
	\centering
	\caption{Reference operating conditions.\strut}
	\label{tab:ref_opcond}
	\scalebox{\tabscale}{\begin{tabular}{llr}
	\toprule
	Symbol & Explanation & Value\\
	\midrule
	$P_\mathrm{A}$ & Gas pressure in anode gas channel & $1.5\,\mathrm{bar}$\\
	$P_\mathrm{C}$ & Gas pressure in cathode gas channel & $1.5\,\mathrm{bar}$\\
	$\mathrm{RH}_\mathrm{A}$ & Relative humidity in anode GC & $100\%$\\
	$\mathrm{RH}_\mathrm{C}$ & Relative humidity in cathode GC & $100\%$\\
	$T_\mathrm{A}$ & Temperature of anode plate and GC & $80\,^\circ\mathrm{C}$\\
	$T_\mathrm{C}$ & Temperature of cathode plate and GC & $80\,^\circ\mathrm{C}$\\
	$\alpha_{\mathrm{O}_2}$ & Oxygen mole fraction in dry oxidant gas & $21\%$\\
	\bottomrule
	\end{tabular}}
\end{table}

In the model parameterization in Sec.~\ref{sec:parameterization}, special attention was paid to six of the most critical and controversial transport parameters for Nafion-based MEAs, and their openly available constitutive relationships were reviewed in Tabs.~\ref{tab:sigma_p} to \ref{tab:ec_rates}. We now employ our macro-homogeneous steady-state MEA model to determine the degree of uncertainty in the fuel cell performance prediction associated with these six material parameterizations. Implemented in COMSOL Multiphysics, the model is numerically solved with the finite element method. 50 elements with quadratic Lagrangian shape functions are used per MEA layer (totaling in 250 finite elements for the entire MEA), and the damped Newton method is used with a relative error tolerance of $10^{-5}$ to solve the coupled nonlinear differential equations simultaneously. Tab.~\ref{tab:ref_opcond} lists the operating conditions at which all simulations shown in this first part are carried out. They were chosen to approximately represent a differential section of a PEMFC operated in an automotive scenario.

\begin{figure*}[t]
	\centering
	\includegraphics{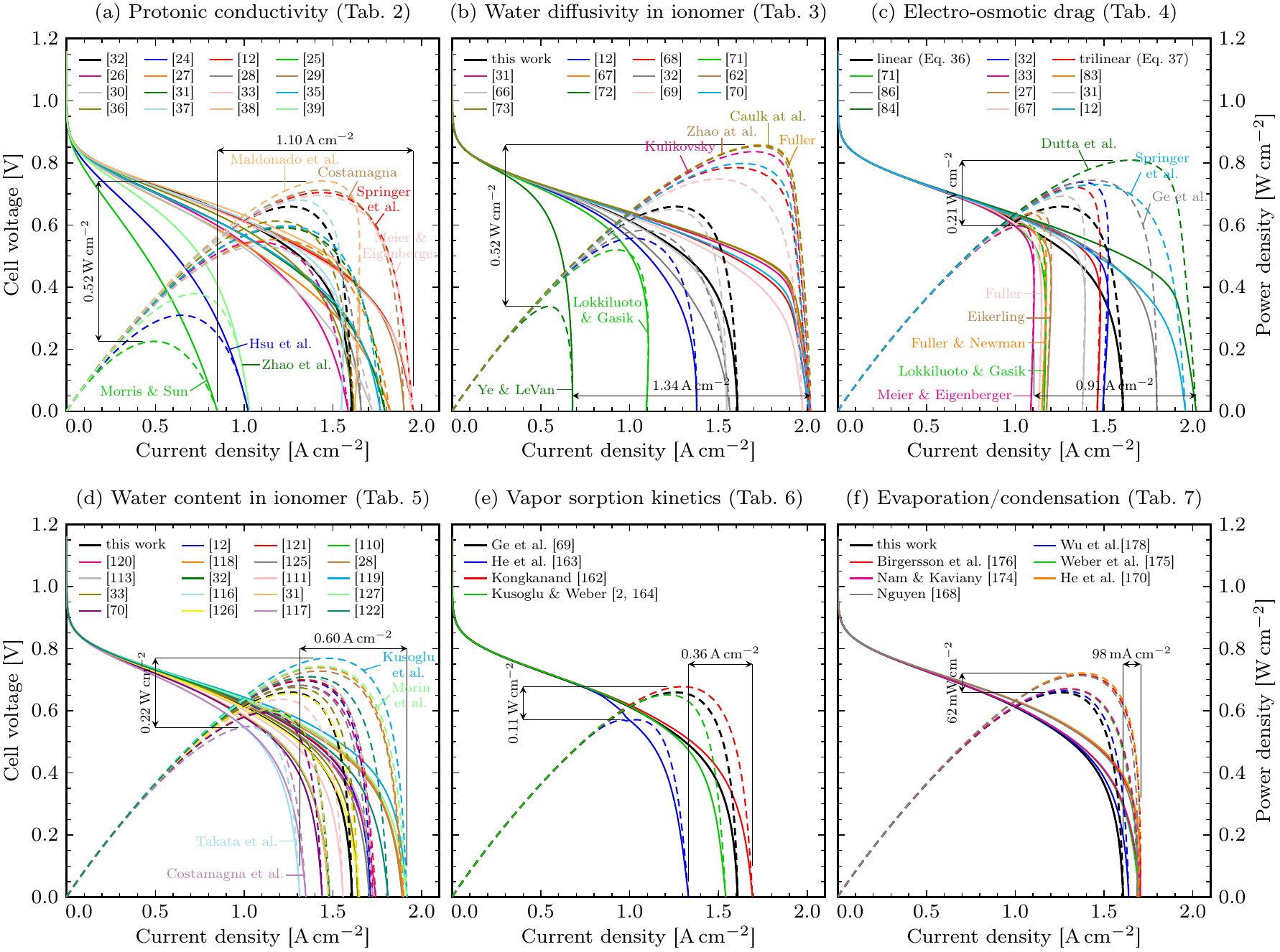}
	\caption{Impact of selected parameterizations on predicted fuel cell performance at reference operating conditions. Solid lines represent voltage (left axes), dashed lines represent power density (right axes). The baseline parameterization is shown with slightly thicker black lines.}
	\label{fig:polarization}
\end{figure*}

\subsection{Protonic conductivity}

Fig.~\ref{fig:polarization} shows the resulting polarization and power density curves predicted by the model when each of the six major parameters are substituted while all others are retained at the baseline as described in Sec.~\ref{sec:parameterization}. As can be recognized from Fig.~\ref{fig:polarization}a, employing different expressions for the ionic conductivity $\sigma_\mathrm{p}$ of the Nafion membrane leads to enormous scatter. In order to quantify the discrepancy between the different model outputs, we selected the maximum reachable current density $I_\mathrm{max}$ and the peak power density $P_\mathrm{max}$ as key figures. From the experimental scatter of $\sigma_\mathrm{p}$ alone, the total bandwidth of obtained values is $1.10\,\mathrm{A\,cm}^{-2}$ for $I_\mathrm{max}$ and $0.52\,\mathrm{W\,cm}^{-2}$ for $P_\mathrm{max}$, making the protonic conductivity of the ionomer a material property of very large uncertainty.

The parameterizations by Hsu et al.~\cite{hsu:80}, Morris \& Sun~\cite{morris:93} and Zhao et al.~\cite{zhao:12}, which were shown to predict generally low conductivities in Fig.~\ref{fig:sigma_p}, yield current densities that are roughly a factor of two poorer than the others, over the entire range of cell voltages. Yet, if these three are dismissed as possible measurement outliers, the remaining performance variation is still appreciable (up to about $0.4\,\mathrm{A\,cm}^{-2}$). To reach large current densities, the ionomer needs to offer high conductivity at low water content, because electro-osmotic drag dries out the anode side of the membrane (cf.~\cite{vetter:18}). It is for this reason that the conductivity data by Maldonado et al.~\cite{maldonado:12} yields high performance at intermediate voltages, followed by a sudden drop. The parameterizations which suggest higher $\sigma_\mathrm{p}$ at low $\lambda$ values (in particular, those by Springer et al.~\cite{springer:91}, Costamagna~\cite{costamagna:01} and Meier \& Eigenberger \cite{meier:04}) outperform others at large current densities. It must be stressed that even small differences in the conductivity expression that may appear marginal at first can have a great impact on fuel cell models and that detailed knowledge of $\sigma_\mathrm{p}(\lambda)$ at low $\lambda$ is critical.

\subsection{Water diffusivity in the ionomer}

Fig.~\ref{fig:polarization}b shows the influence of the water diffusion coefficient $D_\lambda$. Generally, higher diffusivities result in stronger back diffusion, which helps keeping the anode humidified at large current densities, but the effect of nonlinear features in $D_\lambda$ can be subtle. Ye \& LeVan's data \cite{ye:03} yields the poorest cell performance due to the very low diffusivity in the dry and wet regimes, despite being the largest in between (Fig.~\ref{fig:D_lambda}d). The correlation proposed by Lokkiluoto \& Gasik~\cite{lokkiluoto:13}, which globally predicts low diffusivity, also yields a low limiting current density. To maintain high performance at large currents, good water diffusivity under very dry conditions is crucial, similar to our conclusion for the ionic conductivity above. Fuller's \cite{fuller:92b}, Kulikovsky's \cite{kulikovsky:03}, Zhao's \cite{zhao:11} and Caulk's \cite{caulk:12} expressions yield the lowest membrane resistivites because their back diffusion coefficients remain the largest toward low $\lambda$ values (cf.\ Fig.~\ref{fig:D_lambda}). Our own fit \cite{vetter:18} to measurement data by Mittestadt \& Staser \cite{mittelsteadt:11} (with Arrhenius correction using Eq.~\ref{eq:Ed}) lies somewhat in the middle of the overall scatter of polarization curves. With $1.34\,\mathrm{A\,cm}^{-2}$ for $I_\mathrm{max}$ and $0.52\,\mathrm{W\,cm}^{-2}$ for $P_\mathrm{max}$, the uncertainty associated with $D_\lambda$ is even bigger than with $\sigma_\mathrm{p}$.

\subsection{Electro-osmosis}

Next, we turn our attention to the electro-osmotic drag coefficient $\xi$. Generally, the fewer water molecules are dragged along with each traversing proton, the more evenly the membrane remains humidified across its thickness, which in turn reduces the ohmic resistivity of the membrane due to the monotonicity of $\sigma_\mathrm{p}$ in $\lambda$. It thus comes with no surprise that the large variation in measurement data on $\xi$ translates to wide scatter in the resulting polarization curves. As Fig.~\ref{fig:polarization}c shows, the empirical constitutive expressions by Springer et al.~\cite{springer:91}, Dutta et al.~\cite{dutta:01} and Ge et al.~\cite{ge:06} yield the best performance, since they predict the lowest $\xi$ at medium to dry conditions (cf.\ Fig.~\ref{fig:xi}b). It so happens, though, that this is also the regime where the reported experiments diverge the most (cf.\ Fig.~\ref{fig:xi}a). When the parameterizations of Fuller \cite{fuller:92,fuller:92b}, Eikerling \cite{eikerling:98}, Meier \& Eigenberger \cite{meier:04} or Lokkiluoto \& Gasik \cite{lokkiluoto:13} are employed, the fuel cell model stalls at much lower current densities, because the anode dries out more. In summary, the electro-osmotic drag coefficient in Nafion is a model parameter with large uncertainty.

\subsection{Membrane hydration}

A natural implication of the monotonic increase of membrane conductivity with increasing hydration is that higher sorption isotherms yield better performance prediction. This effect is apparent in Fig.~\ref{fig:polarization}d. Takata's \cite{takata:07} and Costamagna's \cite{costamagna:08} vapor sorption models are among those which yield the smallest $\lambda_\mathrm{v}$ values in the low and high activity regimes, respectively (cf.\ Fig.~\ref{fig:lambda_v}). They therefore result in the steepest decline of the polarization curve in the ohmic region. Morin's data \cite{morin:17} and Kusoglu's model, on the other hand, both suggest good membrane hydration over the entire activity range, resulting in a polarization curve that extends toward higher current densities. With a total spread of $0.60\,\mathrm{A\,cm}^{-2}$ for the tested parameterizations, the water vapor uptake of the membrane is a significant source of modeling uncertainty. Given that it depends also on the membrane's hygro-thermal history \cite{zawodzinski:93c,maldonado:12} (which has been ignored in our present analysis), this highlights that detailed experimental characterization of the membrane is required to make PEMFC models predictive.

\subsection{Vapor sorption kinetics}

The next effect in the line is the sorption of water vapor at the ionomer--gas interface. With the four hydration-dependent expressions for the mass transfer coefficients plotted in Fig.~\ref{fig:ad_coeff}, we estimate the modeling uncertainty originating from measurement data on $k_\mathrm{a,d}$, setting $k_\mathrm{a}=k_\mathrm{d}$ for the two parameterizations that do not distinguish between absorption and desorption \cite{he:11,kusoglu:12b}. Fig.~\ref{fig:polarization}e shows the resulting polarization curves, which start to separate no earlier than at intermediate cell voltages. He's expression \cite{he:11} yields the most limited cell performance due to the fast-dropping mass transfer coefficients toward low hydration numbers. The slight kink near 0.6\,V is due to the kink in $k_\mathrm{a,d}$ at $\lambda_0=3.17$, which is used here with the intention of adopting He's expression without modification, in spite of the absence of residual hydration in the employed sorption isotherm (Eq.~\ref{eq:bet}). Konkanand's data \cite{kongkanand:11} yields the largest current densities, for their mass transfer coefficients remain positive (still allowing for moderate vapor absorption) when the electro-osmotic drag dries out the anode side of the membrane. Once again, we find that the constitutive material behavior at very low water content has a major impact in the predicted cell performance and therefore needs to be known with high accuracy -- higher than currently available in the literature.

\subsection{Evaporation and condensation}

Finally, the effect of adopting different expressions for the evaporation and condensation rates is shown in Fig.~\ref{fig:polarization}f. Wu's Hertz--Knudsen expression \cite{wu:09b} is used as the baseline parameterization with lowered evaporation coefficient (see Sec.~\ref{sec:evap_cond}), which results in higher liquid water saturation, and consequently, lower limiting current density. With a difference of just 31\,mA\,cm$^{-2}$, the effect is relatively small, though. Expressions which give fast evaporation and slow condensation, such as \cite{nguyen:99,he:00,birgersson:05} (cf.\ Fig.~\ref{fig:ec_rates}), reduce the liquid water saturation further, in particular in the CCL, allowing for better oxygen access to the catalyst sites. In total, though, the scatter induced by the different phase change rates is much smaller than for the five ionomer properties discussed above, because in the examined parameter range, evaporation is fast enough to yield a relative gas humidity of $\approx100\%$ across almost the entire MEA where liquid water is present.

\subsection{Result summary}

\begin{table}
	\centering
	\caption{Variation in current-voltage characteristics due to scatter in selected parameterizations at reference operating conditions. Parameters are sorted by uncertainty in decreasing order.}
	\label{tab:polarization_variation}
	\scalebox{\tabscale}{\begin{tabular}{lrrrr}
	\toprule
	& \multicolumn{2}{c}{$I_\mathrm{max}$ [A\,cm$^{-2}$]} & \multicolumn{2}{c}{$P_\mathrm{max}$ [W\,cm$^{-2}$]}\\
	\cmidrule(lr){2-3}
	\cmidrule(lr){4-5}
	Parameters & std.~dev. & total spread & std.~dev. & total spread\\
	\midrule
	$D_\lambda$ & 0.43 & 1.34 & 0.16 & 0.52\\
	$\sigma_\mathrm{p}$ & 0.34 & 1.10 & 0.15 & 0.52\\
	$\xi$ & 0.32 & 0.91 & 0.07 & 0.21\\
	$\lambda_\mathrm{v}$ & 0.19 & 0.60 & 0.06 & 0.22\\
	$k_\mathrm{a}$, $k_\mathrm{d}$ & 0.15 & 0.36 & 0.05 & 0.11\\
	$\gamma_\mathrm{c}$, $\gamma_\mathrm{e}$ & 0.04 & 0.10 & 0.03 & 0.06\\
	\bottomrule
	\end{tabular}}
\end{table}

The results of our uncertainty analysis are summarized in Tab.~\ref{tab:polarization_variation}, where in addition to the total spread, also the standard deviations for the limiting current density and peak power density are given. This corroborates that the same ranking list is obtained when individual parameterizations, which may be deemed measurements outliers, are given less weight or omitted from the analysis. Evidently, getting the water transport inside the membrane and across its interfaces correct is the key to predictive macro-homogeneous MEA modeling for PEMFCs. Most uncertainty stems from the ionic conductivity as well as from the two major water transport effects in the ionomer: back diffusion and electro-osmosis.

\section{Conclusion}

With the present work, we have found a quantitative answer to the questions raised in the introductory section: When characterizing a MEA with the intention of extracting material properties for PEMFC modeling, special attention should be paid to the membrane properties, in particular the Fickean diffusivity of dissolved water, the protonic conductivity and the electro-osmotic drag coefficient. At least under the present modeling assumptions and operating conditions, these are the top three traits of Nafion which cause the most uncertainty in the predicted fuel cell performance, based on the available data in the open literature. For high accuracy at large current densities, their functional dependency on the state of membrane hydration is critical, in particular in the dry regime (low $\lambda$). This finding is relevant not only for fuel cell experimentalists, but also for modelers who are faced with the task of selecting appropriate MEA material parameterizations for their calculations.

With the present scatter in the available data records on Nafion properties, predictive performance prediction is very difficult. Measurement errors should be quantified more routinely in experimental studies than they have been in the past. Further research is required to improve the understanding of the transport processes of hydrogen ions and water molecules across the electrolyte membrane, not only under varying water content, but also to clarify the temperature dependence of water uptake and electro-osmotic drag.

In Part I, we have studied only \emph{local} parameter sensitivity, i.e., at fixed operating conditions and with all material parameterizations but one fixed. Given the highly nonlinear nature of PEMFCs, the picture might change in different scenarios. So far, we have focused on uncertainty in the extrema of polarization characteristics, which is induced by scatter in the proposed parameterizations for six of the most critical MEA properties. In Part II, we perform a \emph{global} parameter sensitivity analysis to gain insight into the general model response to changes in the MEA parameters, not only limited to cell performance, but also regarding the predicted heat and water balance.

\section*{Acknowledgements}

We thank Robert Herrend\"{o}rfer for proof reading the manuscript. Funding: This work was supported by the Swiss National Science Foundation [project no.\ 153790, grant no.\ 407040\_153790]; the Swiss Commission for Technology and Innovation [contract no.\ KTI.2014.0115]; the Swiss Federal Office of Energy; and through the Swiss Competence Center for Energy Research (SCCER Mobility).

\end{document}